\documentclass[12pt, preprint]{aastex}
\usepackage{ amsmath, graphicx, fancyhdr, multirow, epsfig, verbatim }%

\newcommand{\ufilter}{\textit{u}}
\newcommand{\gfilter}{\textit{g}}
\newcommand{\rfilter}{\textit{r}}
\newcommand{\ifilter}{\textit{i}\enspace}
\newcommand{\izfilter}{\textit{i}}

\newcommand{\SNgj}{SN~2005gj} 
\newcommand{\SNhc}{SN~2005hc} 
\newcommand{\SNhj}{SN~2005hj} 
\newcommand{\SNir}{SN~2005ir} 
\newcommand{\SNhk}{SN~2005hk} 
\newcommand{\SNku}{SN~2005ku} 
\newcommand{\SNjg}{SN~2007jg} 
\newcommand{\SNjh}{SN~2007jh} 
\newcommand{\SNmm}{SN~2007mm} 

\usepackage{graphicx}
\usepackage{epstopdf}
\usepackage{lscape}

\begin{document}
\title{A Precision Photometric Comparison between SDSS-II and CSP Type Ia Supernova Data}
\author{J. Mosher\altaffilmark{1},
M. Sako\altaffilmark{1},
L. Corlies\altaffilmark{1,2},
G. Folatelli\altaffilmark{3,4},
J. Frieman\altaffilmark{8,6},
J. Holtzman\altaffilmark{5},
S.W. Jha\altaffilmark{7},
R. Kessler\altaffilmark{6,8}, 
J. Marriner\altaffilmark{9},
 M.M. Phillips\altaffilmark{4}, 
 M. Stritzinger\altaffilmark{13,10,11},
 N. Morrell\altaffilmark{4},
  D.P. Schneider\altaffilmark{12}
 }

\email{jmosher@sas.upenn.edu}

\altaffiltext{1}{Department of Physics and Astronomy, University of Pennsylvania, 209 South 33rd Street, Philadelphia, PA 19104, USA}
\altaffiltext{2}{Department of Astronomy, Columbia University, 550 West 120th Street, New York, New York 10027, USA}
\altaffiltext{3}{Institute for the Physics and Mathematics of the Universe (IPMU), University of Tokyo, 5-1-5 Kashiwanoha, Kashiwa, Chiba, 277-8583, Japan}
\altaffiltext{4}{Las Campanas Observatory, Carnegie Observatories, Casilla 601, La Serena, Chile} 
\altaffiltext{5}{Department of Astronomy, MSC 4500, New Mexico State University, P.O. Box 30001, Las Cruces, NM 88003, USA }
\altaffiltext{6}{Department of Astronomy and Astrophysics, University of Chicago, 5640 South Ellis Avenue, Chicago, IL 60637, USA }
\altaffiltext{7}{Department of Physics and Astronomy, Rutgers, the State University of New Jersey, 136 Frelinghuysen Road, Piscataway, NJ, 08854, USA}
\altaffiltext{8}{Kavli Institute for Cosmological Physics, University of Chicago, 5640 South Ellis Avenue, Chicago, IL 60637, USA }
\altaffiltext{9}{Center for Particle Astrophysics, Fermi National Accelerator Laboratory, P.O. Box 500, Batavia, IL 60510, USA}
\altaffiltext{10}{Oskar Klein Centre for Cosmo Particle Physics, AlbaNova University Center, 106 91 Stockholm, Sweden}
\altaffiltext{11}{Dark Cosmology Centre, Niels Bohr Institute, University of Copenhagen, Juliane Maries Vej 30, 2100 Copenhagen \O, Denmark}
\altaffiltext{12}{Department of Astronomy and Astrophysics, The Pennsylvania State University, 525 Davey Laboratory, University Park, PA 16802, USA}
\altaffiltext{13}{Department of Physics and Astronomy, Aarhus University, Ny Munkegade, DK-8000 Aarhus C, Denmark}

\begin{abstract}
 Consistency between Carnegie Supernova Project (CSP) and SDSS-II supernova (SN) survey  
 \ufilter \gfilter \rfilter \ifilter  measurements has been evaluated by comparing SDSS and CSP photometry for nine spectroscopically confirmed Type Ia supernova observed contemporaneously by both programs. The CSP data were transformed into the SDSS photometric system. Sources of systematic uncertainty have been identified, quantified, and shown to be at or below the 0.023 magnitude level in all bands. When all photometry for a given band is combined, we find average magnitude differences of equal to or less than 0.011 magnitudes in $ugri$, with rms scatter ranging from 0.043 to 0.077 magnitudes. The $u$ band agreement is promising, with the caveat that only four of the nine supernovae are well-observed in $u$ and these four exhibit an 0.038 magnitude supernova-to-supernova scatter in this filter. 
\end{abstract}

\section{Introduction}

Used as standard candles, Type Ia supernovae (hereafter SNe~Ia) provided the first direct evidence of cosmic acceleration \citep{Riess:1998,Perlmutter:1999},  and hence the existence of dark energy. 
With cosmic acceleration having been firmly established through both  SNe~Ia 
\citep[e.g.][]{Tonry:2003,Riess:Decel, 2007ApJ...659...98R, AstierSNLS1yr:2006, WVEssence1yr:2007, FreedmanNIRHuDi:2009, 2008ApJ...686..749K, KesslerSDSScosmo:2009, 2010ApJ...716..712A, 2011ApJ...737..102S} and other cosmological measurements, such as the late-time integrated Sach-Wolfe effect \citep{GiannantonioISWCosmo:2008} and X-ray cluster distances \citep{AllenXCCosmo:2008}, sights have now turned to understanding the time-varying nature of  dark energy. 
Distinguishing between competing dark energy theories will require photometric precision of SNe~Ia observations on the 1\% level or better.

Several recent surveys  have discovered and observed more than a  
thousand SNe~Ia extending from intermediate- to high-$z$.
Analysis of  the full set of SNe~Ia indicates that the precision of cosmology measurements is now limited as much by systematic  as by statistical uncertainties 
 \citep[e.g.][]{ Hicken:CfACosmo, KesslerSDSScosmo:2009, Guy:20103yr, Conley:SNLS3yrSYSSERR2011,2011ApJ...737..102S}. 
 Systematic uncertainties are particularly acute in the UV, specifically the observer-frame $U$ band \citep{ KesslerSDSScosmo:2009, 2011ApJ...737..102S}. 
 
 Performing highly acccurate UV observations is a challenging task. The filter response function in this region is difficult to characterize due to the significant role played by the atmosphere in determining its shape on the blue side. Rest-frame UV response functions are more variable from telescope to telescope than in the other bands, making the accurate characterization of the UV filter response function even more important. 
The nearby SN sample suffers from both of these problems, since it is a heterogeneous collection of data taken at many telescopes, and most of the light curve data are reported in the Landolt standard system for which the filter-response functions are not well-defined, particularly in $U$.  For this reason, the SALT2 light curve fitter was not trained with observer frame UV  \citep{Guy:SALT2}. U-band calibration of the nearby sample was identified by \citet{KesslerSDSScosmo:2009} as one of the main sources of the discrepancy between cosmology parameters obtained with the MLCS2k2 and SALT2 light-curve fitting models. Due to these known calibration problems, many analyses, including \citet{KesslerSDSScosmo:2009}, \citet{Conley:SNLS3yrSYSSERR2011} and \citet{2011ApJ...737..102S}, recommend avoiding use of observer-frame UV for fitter training and cosmology.

Significant benefits can be gained from SN~Ia observer-frame UV data if systematic calibration uncertainties can be reduced. Spectral observations at high and low redshifts have shown that the UV portion of the Ia spectrum, particularly below 3500 Angstroms, shows increased diversity compared to the optical \citep{FoleyESSUV:2008, 2008ApJ...674...51E, 2011arXiv1110.5809W,2011ApJ...727L..35C} , even after accounting for extinction \citep{2008ApJ...674...51E}.  Some degree of diversity is expected due to differences in progenitor metallicities \citep{1998ApJ...495..617H, 2000A&A...363..705M,2000ApJ...530..966L,2003ApJ...590L..83T, 2008MNRAS.391.1605S}. However, it is not clear that current metallicity theories can explain the observed range of dispersion \citep{2011ApJ...727L..35C, 2008ApJ...674...51E,2011arXiv1110.5809W}. Progenitor-stellar companion interaction \citep{2010ApJ...708.1025K} and asymmetric explosions \citep{FoleyKasenHighV:2010,  2007ApJ...662..459K} are other possible sources for variations in UV flux.  

Although rest-frame UV photometry is arguably easier to obtain at redshifts of $ z \approx 0.2$ and higher \citep[e.g.][]{AstierSNLS1yr:2006}, low-redshift SN observations allow for the acquisition of a much wider range of ancillary data. Spectropolarimetry and very early and late-time supernova spectra are more easily obtained for low-redshift SNe; these provide valuable information about rise times, progenitor interaction, and explosion asymmetry \citep[see for instance][]{2005ApJ...632..450L, 2012ApJ...744...38F, 2011MNRAS.413.3075M}. Host galaxy metallicity data, especially as pertains to the SN Ia location itself, are also easier to obtain for nearby SNe and can be used to probe the host galaxy-luminosity relationship which has been recently observed by \citet{2010ApJ...715..743K, 2010MNRAS.406..782S, 2010ApJ...722..566L, 2011ApJ...743..172D, 2011ApJ...741..127G}. The ability to link any or all of these spectrum-based measurements to observable features in the UV light curve will be valuable for the interpretation of SN~Ia observations as a function of redshift.
At high redshifts, observed light-curves are limited to the bluer bands in the rest frame. At $z \approx 1.0$, the highest redshift SN observations achievable from the ground, the UV region is critical. Therefore, the interpretation of these high-redshift SNe light curves, or whether they can be used at all for cosmological studies will depend on our understanding of SN~Ia rest frame UV models.

 In this paper, we examine SN~Ia photometry from  the Carnegie Supernova Project \citep[CSP;][]{Hamuy:2006} and the SDSS-II Supernova Survey \citep{FriemanSDSSTECH:2008} which are likely to make up significant fractions of future light-curve training sets. Both of these programs have invested substantial time and effort in characterizing their photometric systems and in ensuring accurate photometry. By comparing data from 10 well-observed SNe~Ia in common, we will examine the consistency of their overall calibrations, particularly in the rest-frame UV. Results of these tests will help determine the viability of observer-frame UV photometry, and the utility of these data in light curve fitter retraining efforts currently underway. 
 
\citet{2010ApJS..190..418G} perform a similar comparison between CfA and LOSS $BVRI$ photometry. Although they found photometry agreement to be reasonable, with mean photometry residuals between 0.01 and 0.02 magnitudes in all bands, their scatter was much larger than expected, ranging from 0.07 to 0.11 magnitudes. It has been postulated by \citet{2011arXiv1107.3555F} that the large scatter is due to $S$-corrections, which were not part of the \citet{2010ApJS..190..418G} analysis.  We will show that by incorporating $S$-corrections, which are not negligible in the redshift range of our overlap objects ($z \in 0.02 - 0.08$), we are able to obtain residual rms scatter of the levels expected, on the order of $\sigma \sim 0.05 $ magnitudes. 
 
 In Section \ref{sec:phot} we present an overview of the CSP and SDSS photometry, describe our $S$-correction technique, and tabulate expected systematic uncertainties. Our magnitude data, including $S$-corrected light curves, and magnitude agreement statistics, are described in Section \ref{sec:Results}. In Section \ref{sec:Discussion} we look at magnitude agreement in each band in more detail, including a discussion of template vs spectrum based $S$-corrections.

\section{Photometry}\label{sec:phot}


Between 2005 and 2007 the CSP low-redshift program and SDSS-II supernova survey observed 16 common objects, of which ten are spectroscopically confirmed Ia supernovae. One of these ten (SN~2006fw) has image registration discrepancies, and has been excluded from our analysis.  We use the remaining nine SN~Ia, listed in Table~\ref{table:SNlist}, as our sample.  In subsections \ref{sec:sdssphot} - \ref{sec:scorr}, we give an overview of how photometry was acquired and how the data were placed on a common system. Accompanying this information, Table~\ref{table:ABoffsets} summarizes the AB offsets used to transform synthetic CSP and SDSS AB magnitudes to their native systems. Subsection \ref{sec:Interp} describes how interpolation was used to transfer SDSS photometry onto the observation dates of the CSP data.  Subsection \ref{sec:scorr-systematics} details the systematic uncertainties expected for this analysis. 

\subsection{SDSS-II Supernova Survey}\label{sec:sdssphot}

The SDSS-II Supernova Survey was one of three main scientific programs carried out by the SDSS-II. Supernova candidates were discovered by repeated imaging scans of a 300 deg$^2$ patch of sky over three fall observing seasons in 2005-7. SDSS-II $ugriz$ \citep{1996AJ....111.1748F} imaging was obtained with the SDSS camera \citep{SDSScamera} attached to the SDSS 2.5 m telescope \citep{1996AJ....111.1748F, SDSStelescope} located at the Apache Point Observatory [APO].  Preliminary photometric processing was carried out at APO \citep{Stoughton:WEBFILT, Tucker:2006}. Photometric zero points for nightly frames were obtained from field stars in the Ivezi\'{c} catalog \citep{Ivezic:2007}. Off-site, supernova candidates were flagged for spectroscopic followup \citep{2008AJ....135..348S}, and SN magnitudes were extracted from images using scene-modeling photometry \citep{SDSSphot} and reported in the SDSS natural magnitude system \citep{Smith:2002, 1999AJ....118.1406L}. For a technical summary of the SDSS see \citet{2000AJ....120.1579Y}; further information can be found in \citet{2001AJ....122.2129H}, \citet{2003AJ....125.1559P} and \citet{2009ApJS..182..543A}.

It has recently been determined that the SDSS-II SN photometry requires declination-dependent corrections to compensate for flat-fielding issues with the PT telescope. These corrections are described in \citet{Betoule:2012}; updated photometry will be released in \citet{Sako:2012}, and is used for this work. 

In combination with the absolute flux standard BD+17$^{\circ}$4708, the SDSS photometric system is defined by its photon-weighted filter response functions, which include SDSS filter, CCD response, telescope transmission and 1.3 standard airmass transmission. Absolute flux calibration, tying SDSS native magnitudes to the AB system,  has been determined using SDSS Photometric Telescope (PT) observations of CALSPEC solar analog stars \citep{Tucker:2006}. This process is described in detail in \citet{SDSSphot}. We have updated the AB offsets to reflect three recent revisions: (1) the February 2010 CALSPEC release, which altered the solar analog SEDs slightly, (2) SDSS 2.5m filter response functions  \citep{Doi:2010}, which apply specifically to observations taken in 2004 or later,  and (3) updated SDSS PT to 2.5m linear magnitude and color transformation equations used to transfer the observed solar analog magnitudes to the SDSS 2.5 meter system. Of these three changes, the filter response function update has the largest effect, particularly on the $u$-band offset which decreases by 0.0316 magnitudes. The changes in $gri$ offsets are at the millimag level or smaller.   

The AB offsets $m_{AB} - m_{SDSS}$ used in this work are $-0.069 \pm 0.005$, $+0.021 \pm 0.004$, $+0.005 \pm 0.004$ and $+0.018 \pm 0.009$ for $ugri$, respectively, where uncertainties have been calculated as per the description in \citet{KesslerSDSScosmo:2009}. These take into account internal consistency of the solar analogs as well as the uncertainty in the central wavelengths of the filter response functions. 

 \subsection{CSP Supernova Program}\label{sec:cspphot}
The CSP optical ($ugriBV$) follow-up campaigns were carried out with
the Direct CCD Camera attached to the  Henrietta Swope 1 m telescope located at 
the   Las Campanas Observatory (LCO).  A subset of field stars from the published Smith catalog\footnote{As the CSP observed with both Sloan and Johnson $B$ and $V$ filters, only stars common to both the Landolt \citep{Landolt:1992} and the Smith catalogs were used for calibrating the photometry of the local sequences.}  were used to calibrate the local sequences used to derive nightly zeropoints. Similarly to the SDSS survey, CSP magnitudes are published in the native photometric system, defined by the SWOPE filter response functions of \citet{Max:2011} and the primary standard BD+17$^{\circ}$4708.

Preliminary throughput curves given by \citet{Hamuy:2006} were updated by \citet{Contreras:2010}, who emphasized that the $u$-band curve remained uncertain. Definitive measurements of the CSP filter throughput curves were carried out at the telescope in 2010 using a monochromator and calibrated photodiodes \citep{2010SPIE.7735E.201R, Max:2011}. We adopt these curves in the present analysis. As with the SDSS, CSP throughput curves include filter transmission, CCD response, telescope transmission, and 1.3 standard airmass transmission.

Absolute flux calibration for the CSP \ufilter\gfilter\rfilter\ifilter photometry is taken from the published \citet{Smith:2002}  magnitudes of the SDSS primary standard star BD+17$^\circ$4708. Color terms are used to transform these magnitudes into the CSP native system. Analogous to the SDSS procedure, AB offsets are determined by comparing expected native CSP magnitudes with synthetic CSP photometry of the CALSPEC SED \verb9bd17d4708_sticsnic_002.ascii9\footnote{The most recent CALSPEC library is available for download at ftp://ftp.stsci.edu/cdbs/current\_calspec/ .}. Using this method, the following $ugri$ offsets were obtained: $-$0.050, $-$0.017, $-$0.005, 0.002. These values are consistent with the zeropoints published in \citet{Contreras:2010}.  We use color term uncertainty as a proxy for the CSP offset uncertainty. Uncertainties are 0.017, 0.009, 0.017 and 0.017 mag for \ufilter\gfilter\rfilter\ifilter \citep{Hamuy:2006}.  A summary of calibration information is provided in Table~\ref{table:ABoffsets}.

\subsection{Calibration Star Comparison}
Magnitudes of calibration stars in common were transformed to the SDSS native system and compared. For the SDSS-II, we chose to use SDSS Data Release 8 \citep[SDSS DR8;][]{2011ApJS..193...29A} ``ubercal'' magnitudes rather than Ivezi\'{c} catalog magnitudes, to match the recalibrated SDSS-II SMP photometry data. Each bandpass of each supernova had between 2 and 14 calibration stars in common. Table~\ref{table:calstars} shows mean and rms calibration star agreement as a function of SN and filter. \SNku ~stands out as having particularly poor agreement, with three of four bands differing by more than 0.06 magnitudes. The $u$ band in general is notable for its lack of agreement, with four of nine SNe having mean calibration star differences of more than 0.06 magnitudes. Rather than exclude poorly agreeing SNe from our already-small data set, we have chosen to combine mean calibration star difference in quadrature with photometric uncertainty.

\subsection{$S$-correction Procedure}\label{sec:scorr}
Before comparing photometry from CSP and SDSS-II  for a given SN~Ia, it is necessary to transform the 
photometry to a common photometric system. This is accomplished through the use of
$S$-corrections  \citep{Stritzinger:2002}. $S$-corrections account specifically for differences between filter response functions and are computed synthetically based on the redshift and spectral energy distribution (SED) of the object. Therefore they require models for both the native photometric 
system of the data and the common system to which we wish to transform to, as well as a reasonable model for the SED. Since SN~Ia SEDs evolve with time, $S$-corrections must be calculated for each observed epoch. Depending on the amount of difference between filters and the underlying SED, $S$-corrections can be minimal to quite significant. As shown in the right-hand panel of Figure~\ref{fig:scorr_mags} , CSP to SDSS $S$-correction magnitudes for a mean SN~Ia observed at a redshift of 0.04 (the typical redshift of our sample) are on the order of $-$0.1 magnitudes for the $u$, +0.05 magnitudes for $i$, and 0.01 magnitudes for $g$ and $r$.  

The $S$-correction technique described by \citet{Suntzeff:2000}, \citet{Stritzinger:2002},  and \citet{Phillips:2005hk} was used to transform CSP photometry to the native SDSS system.  The gist of the technique is to adjust the colors of an appropriate supernova SED until they match the observed colors in the original system, then to use the adjusted template to compute synthetic photometry in the new system. 

Since we do not have observed spectra corresponding to each photometric measurement, Hsiao templates\footnote{http://www.astro.uvic.ca/~hsiao/uber/index.php} \citep{Hsiao:2007},  were used as proxies for the time-evolving SN~Ia SED. The templates were linearly interpolated to the desired rest frame epoch, redshifted, adjusted to the appropriate Milky Way Galaxy extinction using the CCM extinction law \citep{CCMdust:1989} and the dust maps of \citet{dust}, and color-matched to CSP observed photometry for the $x^{th}$ and $(x+1)^{th}$ bands (e.g.,  for \gfilter-band $S$-corrections, the template was color-matched to $g- r$)\footnote{Since CSP optical photometry only extends through $i$, CSP $r$ and $i$-bands were used for \ifilter-band $S$-corrections.}. Following \citet{Nugent:2002}, color-matching was done via the CCM dust extinction law \citep{CCMdust:1989}. For each desired $S$-corection, a Monte Carlo routine was used to determine the variance in synthetic magnitude resulting from observed photometry uncertainties. 

\subsection{Interpolation}\label{sec:Interp} 
We interpolate SDSS light curves to obtain SDSS magnitudes on the dates of CSP photometry. 
Although interpolation with light-curve fitters was considered, ultimately the choice was made to use cubic splines. This decision allowed us to treat Branch-normal and peculiar SNe identically, and avoided potential systematic uncertainties that could be introduced by the set of SN data used to train the light-curve fitter. Since five of our nine SNe are peculiar (see see \S\ref{TemplateMismatch} or Table~\ref{table:SNlist}), the use of a model-independent interpolation technique was especially important for this work. We did use the MLCS2k2 light curve fitter for a limited epoch range near peak $i$ band  where splines had difficulty reproducing the shape of the light curve. To ensure interpolation quality, to be included in our analysis CSP photometry had to be bracketed by SDSS photometry, with at least one of those bracketing points being observed within 2 observer-frame days of the CSP epoch. We also required that interpolated magnitudes have uncertainties less than or equal to 0.05 magnitudes ( 0.06 magnitudes for $u$ ). 

For most of the SNe, CSP and SDSS have very similar sampling rates. Because the rolling SDSS search discovered most of these SNe, interpolating SDSS data allows inclusion of slightly more data near peak. Therefore, seven of the nine supernovae were interpolated from SDSS photometry onto CSP observation dates. For the other two objects, {\SNhc} and {\SNjg}, we interpolated from CSP onto SDSS dates. {\SNhc} was located on the overlap area between the two SDSS-II observing strips where calibrations do not align exactly, leading to difficulties with an SDSS spline interpolation. {\SNjg} was better sampled by the CSP. 

\subsection{Systematic Uncertainties}\label{sec:scorr-systematics}

In this section, systematic uncertainties introduced by calibration, $S$-correction, and interpolation will be discussed and quantified. A summary of the systematic error attributable to each source is given in Table~\ref{table:SYSERR}. 

\subsubsection{Absolute flux calibration and AB offsets}
As described earlier in \S\ref{sec:sdssphot} and \S\ref{sec:cspphot}, SDSS-II calibration is tied to solar analog magnitudes whereas CSP calibration is tied to BD+17$^\circ$4708 . If SDSS used BD+17$^\circ$4708 as its absolute flux calibration, SDSS AB offsets would change to $-$0.0629, 0.0122, 0.0023 0.0144 magnitudes for $ugri$ respectively. The differences between these two sets of offsets, 0.006, 0.008, 0.003 and 0.004 for $ugri$ respectively,   should be considered systematic uncertainties and are listed in row 2 of Table~\ref{table:SYSERR}. We also include the CSP and SDSS AB offset errors, combined in quadrature,  as systematic uncertainties.

\subsubsection{Mismatches Between SED and Template}\label{TemplateMismatch}
The $S$-correction procedure requires the use of a template SN~Ia SED. Templates can introduce a systematic error in two ways. First, if the template features do not reflect average SN~Ia features, the ensemble of $S$-corrections derived from the template may be biased. The Hsiao template used in this work was designed specifically to address this issue. Its features were determined by taking a weighted average of a large number of observed SN spectra, such that it represents a mean SN~Ia; its suitability has been tested for K-corrections from observed frame to the rest frame $B$ band, for the redshifts  0.0 $< z < $ 0.75. Minimal systematic offsets were observed in this band and redshift range, so long as the proper broad-band colors are used to adjust the template \citep{Hsiao:2007}. 

The second way in which a template SED may introduce a systematic error is if it is used with spectroscopically peculiar SNe~Ia. This is a particular concern for our sample, since five of our nine supernovae are spectroscopically peculiar (see Table~\ref{table:SNlist}) . Three of these, {\SNgj} \citep{Aldering:2006, Prieto05gj:2007},  {\SNhj} \citep{Quimby:2005hj},  and {\SNhk} \citep{Phillips:2005hk} have been discussed in the literature.  Furthermore, we have identified two more of our SNe ({\SNjh} and {\SNmm}) as peculiars using available spectra and the photometric criteria detailed in \citet{Krisciunas:2009}.  For {\SNjh} , spectra have features consistent with a 1986G-like object: near peak this SN has a large Si II at 5800 \r{A} to Si II at 6200 \r{A} ratio, characteristic of fast-decliners, but lacks the strong Ti II absorption features seen in 1991bg-like objects. In addition, the $i$ peak date relative to $B_{max}$ and the weak secondary maximum are indicators of a 1986-G type fast-decliner.  Only one very early (rest frame epoch $-$8) spectrum is available for {\SNmm}. A {\tt SNID} fit \citep{SNIDBlondin:2007} of this spectrum agrees with a peculiar classification, with 6 of the top 7 matches belonging either to SN~1999by or SN~1986G. The {\SNmm} light curve lacks a secondary peak in $i$ and its primary maximum falls after the $B_{max}$ date, typical of a 1991bg-like SN.  

Multiple steps have been taken to minimize the effects of actual vs. template SED mismatch.  For the 1986-G and 1991-bg like objects {\SNjh} and {\SNmm}, we use the Nugent 1991bg templates\footnote{Templates may be downloaded from http://supernova.lbl.gov/~nugent/nugent\_templates.html} \citep{Nugent:2002} rather than those of Hsiao. The other four peculiars have multiple observed spectra during the overlap time period. For these objects, in $gri$ bands, we include only those epochs for which observed spectra were available to compute the $S$-corrections. 

Finally, we quantify the systematic and statistical uncertainties in $S$-correction due to SED mismatch by calculating $S$-correction differences between observed spectra and a corresponding set of template spectra. Because we had very few spectra from the Branch-normal SNe in our sample, we used {\tt SNID} \citep{SNIDBlondin:2007} to identify similar SNe for which spectra were publicly available through the SUSPECT Supernova Database\footnote{http://suspect.nhn.ou.edu/~suspect/}. In this manner, a set of 75 spectra from six SNe Ia were chosen as a comparison set for our data sample. A list of comparison SNe and references to their spectra is given in Table~\ref{table:SCOREST}. To be included in this data set, the spectra had to span the rest frame $g$, $r$, and $i$ bands, and have rest frame epochs between -20 and 80 days of peak $B$-band magnitude. Each SUSPECT database spectrum was warped to match the colors of its corresponding Hsiao spectrum, and $S$-corrections and $S$-correction differences were calculated for a redshift of 0.04, the mean redshift of our sample. In a similar manner, a smaller set of 32 spectra from 9 SUSPECT Supernovae was chosen to make $u$-band $S$-correction difference estimates. Because the numbers of spectra with rest-frame $u$-band coverage are small, any SN Ia spectrum with adequate wavelength coverage was used. The list of comparison $u$ SNe is also given in Table~\ref{table:SCOREST}.

Observed spectrum- template spectrum $S$-correction differences were used to calculate mean and rms scatter $S$-correction differences. Mean $S$-correction differences have been included as a systematic uncertainty, and listed in row three of Table~\ref{table:SYSERR}. We find mean differences between template and spectrum-based $S$-corrections to be no greater than 0.005 magnitudes in the $g$,$r$, and $i$ bands.  The mean difference in $u$ is 0.012 magnitudes. 

The rms scatter in $S$-correction differences was found to be 0.055, 0.017, 0.012, and 0.016 for $ugri$, respectively. To properly account for spectrum-template mismatch uncertainties, these values are added in quadrature with the statistical uncertainties of each template $S$-corrected data point.

\subsubsection{ Color-matching technique }\label{colormatch}

As mentioned briefly in \S\ref{sec:scorr}, part of the $S$-correction process is the adjustment of the colors of the SN SED to match the observed SN colors. These color matching techniques have been discussed extensively in K-correction literature, which we summarize here. As with $S$-corrections, SN~Ia K-corrections are commonly computed by using a template to represent the actual supernova SED, and accounting for differences between individual SNe Ia by adjusting the colors of the template to match those computed from broad-band photometry \citep{Nugent:2002}. This procedure works because K-corrections are primarily determined by the shape of the SN continuum. Two main color-adjustment techniques are in use. \citet{Nugent:2002} use the CCM extinction law \citep{CCMdust:1989} to match a single color, spanning the two bands around the spectral region of interest. Others, including \citet{Hsiao:2007}, have suggested that a multi-color adjustment process results in a better match between the template and the actual SED, and therefore yield more accurate K-corrections compared to those made with a single-color adjustment.  

However, if the single color is chosen carefully so that it spans the filter for which the magnitude is needed, there is a minimal difference between the results of the two techniques (see \citealp{Hsiao:2007}; Figure 9). Analysis pipelines such as {\tt SNANA} \citep{SNANA} use the single-color method because it is simpler to implement. In our case, we opt for the single-color CCM adjustment technique to enable the inclusion of $S$-corrections calculated from observed spectra. Very few observed spectra span a wavelength range that permits even a two-color adjustment.  We have implemented a color selection mechanism similar to that described by \citet{KesslerSDSScosmo:2009} to ensure that we are using the best color for a given filter and redshift. Tests have been done to check consistency between our CCM implementation and multi-color adjustment techniques. No differences have been detected for this work; therefore, our choice of color-matching is not a significant source of systematic error.  

\subsubsection{Interpolation}\label{sysinterp}

Based on our relatively small SN sample dominated by peculiars, and on our desire to minimize the possibility of bias in the $u$-band, it was decided to use a spline interpolation to transfer CSP magnitudes to the dates of the SDSS-II observations. 

The main drawback to this choice is the potential for introducing extra scatter in the magnitude difference results. For SNe with relatively large time gaps between data points, or relatively fast changes in light-curve shape, splines may systematically under or over-estimate magnitudes. We expect this effect to be largest in the $i$-band, due to the rapid changes in magnitude associated with the light curve secondary maximum.  MLCS fits to $i$ were considered, but found to also have trouble reproducing fluxes in this region. Since the magnitude of the secondary maximum is not well-correlated with width-luminosity parameters \citep[see for example][]{Folatelli:2010}, it is likely that any current model would have similar difficulties. As a precaution, our data set excluded those CSP epochs for which the nearest SDSS epoch was more than 2 rest-frame days away.

Simulations were used to test the effects of our interpolation scheme on magnitude residuals and scatter. A set of 500 low redshift (z $<=$ 0.08) SNe were simulated with a simple stretch- and color-based spectral model in {\tt SNANA}. Since the main goal of this simulation was to test the impact of the observing cadence on interpolation uncertainties, a reasonable approximation of a SN~Ia light curve was adequate. The simulations were produced with a cadence of 1 observer-frame day. From this pool of ``perfect" SNe light curves, a redshift-weighted "match" was randomly chosen for each SN in our sample (excluding peculiars). The cadences of the observed data were used to create ``SDSS" and ``CSP" sub-sets of the simulated light curve. Finally, the ``SDSS" simulation was splined onto the CSP observation dates, and the interpolations compared to the ``CSP"data. A total of 200 realizations of the data set were obtained and analyzed for each filter. The same cuts were applied to the simulations as to the real data. In all cases, shifts to the mean CSP-SDSS magnitude difference due to our interpolation scheme were less than 0.006 magnitudes. These shifts are reported in the fourth row of Table~\ref{table:SYSERR}, and are included as systematic uncertainties.

\section{Results}\label{sec:Results}

\hspace{0.25in}  

Plots of the light curves of the SNe Ia in our sample are shown in the left panels of Figures ~\ref{fig:SN4524} --~\ref{fig:SN18890}. The plotted magnitudes are all in the SDSS photometric system. Photometric data used in these calculations are compiled in Tables~\ref{table:umagdata}--\ref{table:imagdata}. Magnitude residuals passing quality cuts have been plotted on the right hand panels of Figures ~\ref{fig:SN4524} --~\ref{fig:SN18890} (quality cuts have been described in section \S\ref{sec:Interp}). Two sets of error bars have been used to differentiate between uncertainty with and without calibration star disagreement. It should be noted that only some of the residuals displayed in the right hand panels of Figures ~\ref{fig:SN4524} --~\ref{fig:SN18890} have been included in the data analysis. In particular, none of the template-corrected magnitudes for the spectroscopic peculiars \SNgj, \SNhj, or \SNhk ~were included. These points have been displayed in the figures to illustrate the differences between template and observed spectrum $S$-corrections. Because specialized templates were used for fast-decliners \SNjh ~and \SNmm, their template-corrected magnitudes have been included in the data analysis. 

Two separate methods were used to obtain mean magnitude differences for each filter. First, all data for a given filter was pooled and $ugri$ magnitude residual weighted means and standard deviations were calculated. These results are listed in Table~\ref{table:nopecs_allGaussparams}. Second, individual SN mean magnitude residuals were calculated, then these values were combined. Figure ~\ref{fig:SNscatt} shows the SN-by-SN mean magnitude residuals as a function of filter, and the combined mean magnitude residuals are given in Table~\ref{table:SNscatt}. The first method measures the typical difference and scatter in the difference for a single photometry measurement in a given filter. The second method quantifies the typical difference and scatter in the difference for any one supernova. 

For the $u$, $g$, and $i$ filters, these two methods give consistent agreement estimates. Two of the four mean magnitude differences, $u$ and $g$, are consistent with zero -- indicating agreement between the CSP and SDSS data sets -- at the 1-2$\sigma$ level. From this agreement, we conclude that the likelihood of significant systematic error is small in these bands. The rms error in $g$ is also fairly small: $68 \%$ of individual photometry observations in $g$ agree at better than 0.043 magnitudes, and $68 \%$ of individual SNe will have mean photometry agreeing within $0.028$ magnitudes. The $u$ band scatter is quite a bit larger, with an individual data point having a $68 \%$ chance of agreeing within 0.077 magnitudes. The four SNe in our sample with 3 or more data points have an rms scatter of 0.038 magnitudes. In the $i$ band, we observe an overall systematic offset of -0.011 magnitudes, inconsistent with zero at the 2.2 $\sigma$ level. The scatter around this value is slightly larger than in $g$, with an point-by-point rms of $0.050$ magnitudes, and a SN by SN rms of $0.032$ magnitudes. Finally, the $r$ band gives slightly inconsistent results between point-by-point or SN-by-SN measurements. Based on the SN-by-SN measurements shown in Table~\ref{table:SNscatt} and Figure ~\ref{fig:SNscatt}, it appears that there is a slight systematic difference of 0.011 magnitudes between CSP and SDSS observations in this filter. The point-by-point calculation is pulled lower by some severe outlier points in \SNmm, which will be discussed in more detail in section \S\ref{sec:outliers}. 

Absolute magnitude differences in all bands, including $u$, are at or below 0.012 mag, regardless of the method used to compute them.  In all cases magnitude differences are comparable to or smaller than the systematic uncertainties listed in Table~\ref{table:SYSERR}. The rms mean scatter scales roughly as one would expect, given uncertainties from photometry, interpolation, and template mismatch.  For the pooled filter averages, we also calculate the scatter in units of the error \footnote{To be precise, for the weighted average we calculate a chi-squared value rather than the bias standard deviation, but functionally they amount to the same thing. }. If our uncertainties are gaussian and correctly estimated, bias should follow a gaussian distribution, and should therefore have a standard deviation consistent with one. For the $u$ and $g$ bands, this is the case, indicating that the uncertainties are reasonable. The $r$ and $i$ band show slightly larger scatter, but in general are consistent. 

To study our residuals in greater detail, we plot them as a function of phase, as shown in the left hand panels of Figure ~\ref{fig:phaseplot}. For the $u$ and $g$ bands, outliers are randomly distributed among SNe. For instance, in the $g$ band, there are 5 data points with residual magnitudes larger than $\pm$ 0.1, belonging to four different supernovae. The phases of these points are fairly evenly distributed across the observed range of -10 to 40 rest frame days. This is less the case with the $r$  band where four of the five residuals larger than 0.08 mags belong to the same supernova, \SNjg, and are at predominantly late times, phases larger than 30. 
In $i$, there appears to be a slight trend in the residual as a function of phase. A closer look at the individual SNe reveals that \SNhc, \SNhk, \SNjg, and \SNmm all show signs of residual magnitude increasing as a function of phase, particularly between phases -10 to 5. However, a fit to the residual shows no significant correlation. 
The right hand panels of Figure ~\ref{fig:phaseplot} demonstrate that all residual distributions are reasonably gaussian.

\section{Discussion and Conclusions}\label{sec:Discussion}

In this work we have chosen to make a direct comparison of the absolute flux calibrations of the CSP and SDSS supernova surveys. This comparison is particularly interesting with respect to the observer frame $u$ band, where such direct absolute flux measurements are rare, and the potential impact on cosmology results is significant. By opting for a quantitative comparison of actual SN Ia observations, we are including all possible effects that could influence agreement of SN Ia flux measurements: calibration differences, pipeline differences, S-correction differences, and template selection. These are the very same effects that will be present when this data is used for light curve fitting or light curve training. In this sense the comparison is more realistic than a calibration star analysis. On the other hand, the sample size available to us is very small (especially in the $u$-band), and the results of all of these effects are mingled. This makes it difficult for us to attribute the observed scatter in $u$ ($0.038$ magnitudes from supernova to supernova) to any single source. 

We can speculate as to the origins of the scatter. For instance, the well-observed Branch-normal \SNhc ~was located on the overlap between the SDSS-II N and S data strips, and shows a systematic offset in SDSS magnitudes between these two sets of data. \SNjg, also a Branch-normal, had a large gap in SDSS observations near peak, and may be more likely than the rest to suffer from interpolation errors. The remaining two SNe, \SNhk ~and \SNhj, are both spectroscopic peculiars for which no observed $u$-band spectra are available to evaluate the suitability of the template. However, to convincingly disentangle these effects would require either a larger sample size or a selection of observed spectra for each supernova spanning at least the 3000 to 6000 Angstrom wavelength range in the supernova rest frame. 

In the $gri$ bands for which more data and more observed spectra are available, agreement is more convincing. As shown in Figure ~\ref{fig:SNscatt}, the majority of the SNe cluster at similar magnitude offsets in each filter, well within the expected systematic uncertainty limits we have quantified in  Table~\ref{table:SYSERR}. Even so, there are one or two outliers in each band which merit discussion. 

\subsection{Outlier SNe in $gri$}\label{sec:outliers}

As Figure ~\ref{fig:SNscatt} makes clear, several SNe have photometry which disagree badly in the $g$, $r$, or $i$-bands. The object \SNmm ~has a mean magnitude difference of approximately -0.07 mags in both  $g$ and $r$. With the exception of a ten-day gap in SDSS $g$-band coverage, it is well-enough sampled by both groups that interpolation errors should not be a problem. Based on a single early-time spectrum and the overall color evolution of this object, we have tentatively classified it as 1991-bg like, and used the Nugent 1991bg templates for its $S$-corrections. No observed spectra were available with which to check the $S$-corrections. The other fast-decliner in our sample, \SNjh, is an outlier in $i$. Unlike \SNmm, \SNjh ~does have a large gap in SDSS observations around peak B-band maximum which could possibly impact interpolated data points at the edges of the gap. This appears to be the case for the $g$-band point at rest-frame epoch 13, where both the template and the observed-spectrum $S$-corrected CSP photometry disagree with their SDSS counterpart.  However, problems in $i$ are not solely due to interpolation difficulties. As shown  in the left-hand panel of Figure~\ref{fig:SN17786}, CSP and SDSS both observed this object near rest frame epochs -1.5 and 25.3, and in neither case is a good agreement between $S$-corrected CSP photometry and SDSS photometry obtained. Although these two objects are the faintest in our data set, the pattern of magnitude differences observed is inconsistent with CSP-SDSS galaxy subtraction differences. Therefore, we conclude that the most likely cause for the observed $g$ and $r$ discrepancy with \SNmm ~and $i$ discrepancy with \SNjh ~is template mismatch. 

Another outlier seen in the $i$-band is \SNjg. As with \SNjh, there are several points observed simultaneously by both groups whose photometry differs even after $S$-correction, suggesting that interpolation is not the source of the discrepancy. The good agreement between \SNjg ~photometry when an observed spectrum is used suggests that the observed $i$ disagreement results from template mismatch. Finally, some mention should be given to \SNir ~which shows quite a large mean magnitude disagreement in $g$-band. This supernova has the highest redshift of any of our sample, and is situated near the core of its host galaxy.The pattern of magnitude disagreement with rest frame epoch is suggestive of a galaxy subtraction difference between the two groups. 

\subsection{Stellar calibration and \SNhc}

In order to achieve a somewhat normal distribution of outliers, it was necessary to take into consideration the individual objects' calibration star differences.  Calibration star discrepancies were largest in $u$, but were instrumental in reducing outliers  in $g$ and  $i$ as well.  \SNku ~was found to have especially poor calibration star agreement. A full analysis of the calibration stars is beyond the scope of this work. A joint effort between the SDSS and SNLS collaborations includes an in-depth analysis of the those two surveys' inter-calibration, and will address in detail the discrepancies observed in the SDSS photometry and the fixes that have been deployed. 

As mentioned earlier in this work, the SDSS-II Supernova Survey suffers from flat-fielding issues which have necessitated declination-dependent corrections to the photometry. Particularly affected by these flat-fielding problems were objects located on the overlap between the SDSS-II N and S data strips such as \SNhc.  This Branch-normal object was very well-observed by both the SDSS and the CSP, makes up a large percentage of our data set, and continues to be over-represented in individual two and three sigma outlier data points, particularly in the $u$ and $i$ bands. 

The effects of the combination of photometry taken on two separate CCD's on alternating observation passes can be seen in Figure ~\ref{fig:SN5944}. The SDSS photometry for this object shows a stair-step effect in all four bands, with the magnitude difference between the two sets of observations varying as a function of epoch. It appears that one set of observations agrees better with the CSP measurements than the other. Since this object makes up 25\% of our data sample, we chose not to eliminate it, and minimized the effects of the offset by interpolating the CSP rather than SDSS data. The calculation of $S$-correction difference information from observed spectra requires interpolating both sets of photometry. Thus, interpolation difficulties with the SDSS data are the likely cause of the large discrepancy of the single observed spectrum data point in $i$.  

\subsection{Conclusions}

Using SN Ia photometry, spectra, and templates, we have checked the consistency of CSP and SDSS SN Ia data. Overall, our analysis gives results well in line with expectations: in $gri$ bands, we obtain photometry agreement at or below the 1\% level in flux with typical epoch-to-epoch scatter no greater than 0.05 magnitudes. These results serve as a sanity check on our comparison technique. In the $u$-band,  we also find observations of the CSP and SDSS to be consistent, and to agree to better than 1\% in flux.  At 0.077 magnitudes, the rms scatter on individual observations is larger than in the $gri$ bands, but is consistent with the correspondingly larger template - spectrum $S$-correction uncertainty.  However, at 0.038 magnitudes, supernova to supernova scatter is fairly large in the $u$ band and our sample size is small, making it difficult to disentangle calibration, pipeline, and $S$-correction differences. Applying a conservative interpretation, we conclude that systematic offsets in observer frame $u$ are equal to or less than 0.04 magnitudes, smaller than the uncertainties currently being added to light curve fitters such as SALT2, and a promising result for ground-based observer frame $u$. 

Through simulations and use of catalog spectra, we were able to quantify the biases introduced into our analysis by interpolation and the use of templates for $S$-corrections. These biases were found to be small in comparison with calibration systematics. Uncertainties due to template use were also estimated and included in our error models. Based on a selection of SUSPECT database spectra chosen to match our sample, we found that template-spectrum mismatch was much higher for the $u$ band (0.055 magnitudes) than in $gr$ and $i$ ( 0.012 - 0.017 magnitudes).  Therefore, more observed spectra covering the rest-frame $u$ band would be required to improve this measurement. 

Finally, two key points should be emphasized.  First, our data set was small, and we were therefore obliged to include SNe that we might otherwise have chosen to cut. Five of the nine SNe in our sample are spectroscopic peculiars. For three of these, \SNgj, \SNhj, and \SNhk, we had at least some usable observed spectra, and we chose to use only this data for our $gri$ sample. For others, particularly the fast-decliners \SNjh ~and \SNmm, we made use of specialized templates. None of the available observed spectra covered the rest frame $u$ band. As a result, all of our $u$-band $S$-corrections rely on color-corrected templates. A sixth object in our data set, \SNhc, sat on the overlap between the SDSS-II N and S data strips, and shows a systematic offset in magnitudes between these two sets of data. Second, by using a direct comparison of supernova observations to calculate the absolute flux calibration of these two surveys, we are tacitly comparing two separate pipelines, with different photometric methods and different host galaxy subtractions. These confounding factors may be muddying the picture.

  \section{Acknowledgements}
  
  J.M. wishes to thank Chris D'Andrea, Ravi Gupta,  and John Fischer for helpful discussions regarding analysis and image reduction, and Ryan Foley for advice on throughput function measurement. 
  Funding for the SDSS and SDSS-II has been provided by the Alfred P. Sloan Foundation, the Participating Institutions, the National Science Foundation, the U.S. Department of Energy, the National Aeronautics and Space Administration, the Japanese Monbukagakusho, the Max Planck Society, and the Higher Education Funding Council for England. The SDSS Web Site is http://www.sdss.org/.
  The SDSS is managed by the Astrophysical Research Consortium for the Participating Institutions. The Participating Institutions are the American Museum of Natural History, Astrophysical Institute Potsdam, University of Basel, University of Cambridge, Case Western Reserve University, University of Chicago, Drexel University, Fermilab, the Institute for Advanced Study, the Japan Participation Group, Johns Hopkins University, the Joint Institute for Nuclear Astrophysics, the Kavli Institute for Particle Astrophysics and Cosmology, the Korean Scientist Group, the Chinese Academy of Sciences (LAMOST), Los Alamos National Laboratory, the Max-Planck-Institute for Astronomy (MPIA), the Max-Planck-Institute for Astrophysics (MPA), New Mexico State University, Ohio State University, University of Pittsburgh, University of Portsmouth, Princeton University, the United States Naval Observatory, and the University of Washington. 
  Support for this research at Rutgers University was provided in part by NSF CAREER award AST-0847157 to SWJ. 
  
  This work is based in part on observations made at the following telescopes. 
  The APO 3.5 m telescope is owned and operated by the ARC. We thank the observatory director, Suzanne Hawley, and site manager, Bruce Gillespie, for their support of this project. The Subaru Telescope is operated by the National Astronomical Observatory of Japan. The William Herschel Telescope is operated by the Isaac Newton Group on the island of La Palma in the Spanish Observatorio del Roque de los Muchachos of the Instituto de Astrofisica de Canarias. Observations at the ESO New Technology Telescope at La Silla Observatory were made under programme IDs 77.A-0437, 78.A-0325, and 79.A-0715 . Kitt Peak National Observatory, National Optical Astronomy Observatories (NOAO), is operated by the Association of Universities for Research in Astronomy, Inc. (AURA) under cooperative agreement with the NSF. The WIYN Observatory is a joint facility of the University of Wisconsin-Madison, Indiana University, Yale University, and NOAO. The W. M. Keck Observatory is operated as a scientific partnership among the California Institute of Technology, the University of California, and NASA. The Observatory was made possible by the generous financial support of the W. M. Keck Foundation. The South African Large Telescope of the South African Astronomical Observatory is operated by a partnership between the National Research Foundation of South Africa, Nicolaus Copernicus Astronomical Center of the Polish Academy of Sciences, the Hobby-Eberly Telescope Board, Rutgers University, Georg-August-UniversitŠt Gšttingen, University of Wisconsin-Madison, University of Canterbury, University of North Carolina-Chapel Hill, Dartmouth College, Carnegie Mellon University, and the United Kingdom SALT consortium. A.V.F.'s supernova group at U.C. Berkeley is supported by NSF grant AST-0607485.
  
Thanks also to the SUSPECT Online Supernova Spectrum Archive. 

  \clearpage
 
\bibliographystyle{apj}
\bibliography{myrefs}

  \clearpage
 
 \begin{center}
\begin{deluxetable}{cccccccc}
\tabletypesize{\normalsize}
\tablecolumns{8}
\tablewidth{0pt}
\tablecaption{Spectroscopically confirmed overlap SNe Ia \label{table:SNlist}}
\tablehead{
  \colhead{SDSS-II ID} &
  \colhead{SN IAU name} &
  \colhead{z\tablenotemark{a}} &
  \colhead{$B_{max}$, MJD} &
  \colhead{$\Delta m_{15}$} & 
  \colhead{MWG $A_v$\tablenotemark{b}} &
  \colhead{CSP phot. version} &
  \colhead{peculiar}
  }
\startdata
 4524 & 2005gj & 0.0616 & 53658.0 & $\cdots$ & 0.312 & Prieto et al. & yes\tablenotemark{c} \\
 5944 & 2005hc & 0.0459 & 53666.6 & 0.85 & 0.092 & Contreras et al. &       \\
 6558 & 2005hj & 0.0574 & 53673.9 & 0.72 & 0.121 & Stritzinger et al. & yes\tablenotemark{d}  \\
 7876 & 2005ir & 0.0764 & 53684.3 & 0.84 & 0.095 & Contreras et al. &  \\
 8151 & 2005hk & 0.0131 & 53684.8 & $\cdots$ & 0.077 & Phillips et al. & yes\tablenotemark{e} \\
 10805 & 2005ku & 0.0455 & 53697.7 & 1.02 & 0.095 & Stritzinger et al. & \\
 17784 & 2007jg & 0.0371 & 54367.0 & 1.17 & 0.330 & Stritzinger et al. & \\
 17886 & 2007jh & 0.0401 & 54366.0 & 1.77 & 0.321 & Stritzinger et al.  & yes\tablenotemark{f}\\
 18890 & 2007mm & 0.0664 & 54392.2 & 1.91 & 0.113 & Stritzinger et al. & yes\tablenotemark{g}\\
\enddata
 \tablenotetext{a}{Redshifts are in heliocentric frame.}
 \tablenotetext{b}{Taken from \citet*{dust} dust maps.}
 \tablenotetext{c}{SN~2002ic-like.} 
 \tablenotetext{d}{SN~2005-hj like.} 
 \tablenotetext{e}{SN~2002cx-like.} 
 \tablenotetext{f}{SN~1986-G like.} 
 \tablenotetext{g}{SN~1991-bg like.} 
 \end{deluxetable}
 \end{center}

\clearpage
   
\begin{center}
\begin{deluxetable}{cccccc}
\tablewidth{0pt}
\tabletypesize{\normalsize}
\tablecaption{AB offsets for the SDSS and CSP photometric systems \label{table:ABoffsets}}
\tablehead{
	\colhead{}&
	\colhead{\ufilter} &
	\colhead{\gfilter} &
	\colhead{\rfilter} &
	\colhead{\izfilter}
	}
\startdata
  SDSS & $-$0.069 $\pm$ 0.005  & 0.021 $\pm$  0.004 & 0.005 $\pm$ 0.004 & 0.018 $\pm$ 0.009 \\
  CSP  & $-$0.050 $\pm$ 0.017  & $-$0.017 $\pm$ 0.009 & $-$0.005 $\pm$ 0.017 &  0.002 $\pm$ 0.017\\
\enddata
\tablecomments{As pointed out in \citet{SDSSphot}, the SDSS AB offsets are derived by comparing native and synthetic AB magnitudes of the solar analog stars P330E, P177D, and P041C. CSP AB offsets are obtained by comparing native and synthetic AB magnitudes of the CALSPEC standard BD+17$^\circ$4708, and are consistent with zeropoints published in \citet{Contreras:2010}.  }
\end{deluxetable}
\end{center}

\clearpage 

 \begin{center}
\begin{deluxetable}{lcccc}
\tablewidth{0pt}
\tabletypesize{\normalsize}
\tablecaption{CSP SDSS-II Calibration Star Comparison\label{table:calstars}}
\tablehead{
        \colhead{SN}&
        \colhead{$avg \Delta u$} &
        \colhead{$avg \Delta g$} &
        \colhead{$avg \Delta r$} &
        \colhead{$avg \Delta i$}
        }
\startdata
\SNgj & 0.014(49), 9 & -0.005(31), 13 & -0.008(33), 13  & -0.016(57), 13  \\
\SNhc & -0.041(12), 2 & -0.065(104), 3 & -0.028(52), 3 & -0.010(40), 3 \\
\SNhj &  -0.093(119), 4 &  0.024(11), 8 & 0.009(12), 8 & 0.024(42), 8  \\
\SNir &  -0.014(58), 7 & 0.019(30), 14 & 0.015(14), 14 & 0.017(16), 14 \\  
\SNhk &  -0.005(46), 12 & 0.018(15), 15 & 0.019(10), 15 & 0.019(14), 15 \\  
\SNku & 0.212(764), 3 & -0.017(297), 7 & 0.070(105), 7 & 0.085(78), 7 \\
\SNjg &  0.018(58), 8 &  -0.012(26), 11 & -0.016(17), 11 & -0.021(66), 11  \\
\SNjh & 0.069(68), 8  & -0.037(20), 11 & -0.019(20), 11  & -0.003(16), 11  \\
\SNmm & 0.093(18), 4  & 0.009(46), 8 & -0.0002(40), 8  & 0.005(61), 8  \\
\enddata
\parbox{4in}{\tablecomments{ CSP stellar magnitudes have been transformed into the SDSS-II photometric system,  compared with the SDSS-II magnitudes,  and agreement statistics for each SN calculated. The numbers in parentheses are the rms differences given in units of thousandths of magnitudes, followed by the number of stars in common for the listed SN and band. \SNku  ~ stands out as having especially poor agreement,  with CSP calibration stars appearing dimmer than SDSS calibration stars by 0.2 magnitudes in the $u$ bandpass. Based on this evidence, \SNku~ has been omitted from our analysis .  }}
\end{deluxetable}
\end{center}

\clearpage

\begin{center}
\begin{deluxetable}{lllll}
\tablewidth{0pt}
\tabletypesize{\normalsize}
\tablecaption{Systematic Errors affecting synthetic CSP magnitudes\label{table:SYSERR}}
\tablehead{
	\colhead{source of uncertainty}&
	\colhead{$u$} &
	\colhead{$g$} &
	\colhead{$r$} &
	\colhead{$i$}
	}
\startdata
AB Offset Uncertainties &  0.018 & 0.010 & 0.017 & 0.019 \\  
SDSS Absolute Flux Calibration & 0.006 & 0.009 & 0.003 & 0.004 \\
S-Correction Template &  0.012 &  0.005 & 0.000 & 0.001  \\
Interpolation  & 0.006  & 0.001 & 0.005  & 0.002  \\
\hline
Total & $0.023$ & $ 0.014 $ & $0.018$ & $0.020$ 
\enddata
\parbox{4in}{\tablecomments{This table summarizes sources and magnitudes of systematic errors introduced by placing CSP and SDSS photometry on a single system. In the first row, CSP and SDSS AB offset uncertainties have been combined by addition in quadrature. Row two shows the difference in SDSS magnitude that would be obtained were the absolute flux calibration to be tied to BD+17$^{\circ}$4708 rather than Solar Analogs. Row three gives estimates of magnitude differences due to the use of templates rather than spectra in S-correction calculations. Row four gives uncertainty in the mean due to interpolation biases. For more information, see sections \S\ref{sec:sdssphot}, \S\ref{sec:cspphot}, and  \S\ref{sec:scorr-systematics}.}}
\end{deluxetable}
\end{center}
 
 \clearpage
 
 \begin{center}
\begin{deluxetable}{cccc}
\tabletypesize{\normalsize}
\tablecolumns{4}
\tablewidth{0pt}
\tablecaption{SN Ia Used For $S$-Correction Uncertainty Estimation \label{table:SCOREST}}
\tablehead{
  \colhead{SN IAU name} &
  \colhead{peculiar} &
  \colhead{reference} & 
  \colhead{phases used}
}
\startdata
\sidehead{$u$-band comparison}
1960R & $\cdot$ & \citet{2000PASP..112.1439B} & 29 \\
1981B & HVG & \citet{1983ApJ...270..123B} & 0 \\
1994D & $\cdot$ & \citet{1996MNRAS.278..111P} & $-$11, $-$4, 24 \\
1996X & $\cdot$ & \citet{2001MNRAS.321..254S} & 0, 1, 7, 56, 57, 87 \\
1999ee & $\cdot$ & \citet{2002AJ....124..417H} & $-$11 \\
2002bo & HVG & \citet{2004MNRAS.348..261B} & 4, 43 \\
2004dt & HVG & \citet{2007AandA...475..585A} & $-$9, $-$7, $-$6, $-$4, 2, 3, 4, 10, 14 \\
2004eo & IVG & \citet{2007MNRAS.377.1531P} & $-$3, 2 \\
2005cf & $\cdot$ & \citet{2007MNRAS.376.1301P} & $-$8, $-$7, $-$6, $-$3, $-$2, $-$1, 5 \\
\sidehead{$gri$-band comparison}
1994D & $\cdot$ & \citet{1996MNRAS.278..111P} & $-$5, $-$4, $-$2, 2, 4, 10, 11, 24, 26, 50, 76\\
1998bu & $\cdot$ & \citet{2001ApJ...549L.215C} & 10 \\
2002bo & HVG & \citet{2004MNRAS.348..261B} & $-$4, $-$3, $-$2, $-$1, 4, 28, 38 \\
2002er & $\cdot$ & \citet{2005AandA...436.1021K} & $-$8 - 0, 2, 4, 5, 10, 12, 13, 16, 17,20, 34 \\
2003cg & $\cdot$ & \citet{2006MNRAS.369.1880E} & $-$8, $-$7, $-$6, $-$5, $-$1, 1, 12, 16,26, 43 \\
2003du & $\cdot$ & \citet{2007AandA...469..645S} & $-$11, $-$7, $-$5, $-$3, $-$2, 1, 2, 3,4, 6, \\
       &         &                               & 9, 10, 13, 17, 21, 24, 34, 37, 39, 51, 63, 72 \\

\enddata
\end{deluxetable}
\end{center}

 \clearpage
 

\begin{center}
\begin{deluxetable}{lllcc}
\tablecolumns{5}
\tabletypesize{\footnotesize}
\tablewidth{0pt}
\tablecaption{Magnitude agreement statistics: pooled data \label{table:nopecs_allGaussparams}}
\tablehead{
        \multicolumn{2}{c}{} &
        \multicolumn{3}{c}{residual}\\ 
        \hline\\[0.5pt]
	\colhead{band} &
        \colhead{N} &
        \colhead{mean [mags]} &
        \colhead{scatter[mags]} & 
        \colhead{scatter[$\sigma$]}
        }
\startdata
$u$  & 32 & 0.001 $\pm$ 0.014 & 0.077 & 1.01\\
                     \\
$g$ & 62 & -0.002 $\pm$ 0.006 & 0.043 & 0.97\\ 
                     \\
$r$ & 60 & -0.002 $\pm$ 0.005 & 0.049 & 1.24\\ 
                           \\
$i$ & 59 & -0.011 $\pm$ 0.005 & 0.050 & 1.32\\ 
\\
\hline
\enddata
\tablecomments{ Residual is defined as CSP magnitude minus interpolated SDSS magnitude. CSP magnitudes have been $S$-corrected onto the SDSS photometric system. Residual mean and scatter have been calculated using the inverse variance as weight. To test gaussianity of the statistical errors, we have also calculated the scatter in units of the error  $\equiv \Delta m/\delta m$. If errors are random, we expect this quantity to be 1. }
\end{deluxetable}
\end{center}

 \clearpage
 

\begin{center}
\begin{deluxetable}{lllc}
\tablecolumns{4}
\tabletypesize{\footnotesize}
\tablewidth{0pt}
\tablecaption{Magnitude agreement statistics: SN data \label{table:SNscatt}}
\tablehead{
        \multicolumn{2}{c}{} &
        \multicolumn{2}{c}{residual}\\ 
        \hline\\[0.5pt]
	\colhead{band} &
        \colhead{SNe} &
        \colhead{mean [mags]} &
        \colhead{scatter[mags]} 
        }
\startdata
$u$  & 4 & -0.008 $\pm$ 0.016 & 0.038\\

$g$ & 7 & -0.002 $\pm$ 0.006 & 0.028\\

$r$ & 6 & 0.011 $\pm$ 0.005 & 0.025\\
                           
$i$ & 7 & -0.012 $\pm$ 0.005 & 0.032\\
\hline
\enddata
\tablecomments{ Residual is defined as CSP magnitude minus interpolated SDSS magnitude. CSP magnitudes have been $S$-corrected onto the SDSS photometric system. Residual mean and scatter have been calculated using the inverse variance as weight. }
\end{deluxetable}
\end{center}

  \clearpage
 

\begin{center}
\begin{deluxetable}{llllllll}
\tablewidth{0pt}
\tabletypesize{\tiny}
\tablecaption{Magnitude data - $u$ band \label{table:umagdata}}
\tablehead{
        \colhead{IAU} &
        \colhead{MJD}&
        \colhead{CSP(SDSS)} &
        \colhead{SDSS(native)}
        }
\startdata
2005hc   &      53663.3   &      17.9470(0.020)   &      17.9860(0.029)  \\
2005hc   &      53664.4   &      17.9538(0.017)   &      17.9470(0.035)  \\
2005hc   &      53665.4   &      17.9856(0.013)   &      18.0170(0.027)  \\
2005hc   &      53666.4   &      17.9823(0.019)   &      17.9590(0.022)  \\
2005hc   &      53668.3   &      18.0659(0.015)   &      18.0940(0.021)  \\
2005hc   &      53669.3   &      18.1445(0.013)   &      18.1950(0.026)  \\
2005hc   &      53669.4   &      18.1535(0.013)   &      18.1920(0.022)  \\
2005hc   &      53670.3   &      18.2385(0.013)   &      18.2250(0.025)  \\
2005hc   &      53671.4   &      18.3337(0.017)   &      18.4010(0.045)  \\
2005hc   &      53673.3   &      18.4868(0.017)   &      18.4970(0.032)  \\
2005hc   &      53674.3   &      18.5788(0.014)   &      18.6490(0.024)  \\
2005hc   &      53675.3   &      18.6767(0.019)   &      18.7450(0.027)  \\
2005hc   &      53676.4   &      18.7829(0.028)   &      19.0670(0.037)  \\
2005ir   &      53684.2   &      19.1805(0.039)   &      19.2437(0.056)  \\
2007jg   &      54363.3   &      18.7860(0.026)   &      18.7620(0.025)  \\
2007jg   &      54364.4   &      18.7633(0.028)   &      18.7057(0.028)  \\
2007jg   &      54376.3   &      19.8565(0.051)   &      19.7794(0.051)  \\
2005hj   &      53671.4   &      18.1984(0.023)   &      18.2260(0.029)  \\
2005hj   &      53674.3   &      18.2949(0.020)   &      18.3160(0.023)  \\
2005hj   &      53676.4   &      18.4235(0.044)   &      18.5190(0.026)  \\
2005hk   &      53675.1   &      16.8820(0.013)   &      16.8131(0.015)  \\
2005hk   &      53677.1   &      16.5532(0.012)   &      16.5634(0.015)  \\
2005hk   &      53682.1   &      16.4040(0.012)   &      16.4085(0.032)  \\
2005hk   &      53683.1   &      16.4199(0.015)   &      16.4274(0.029)  \\
2005hk   &      53684.1   &      16.4741(0.015)   &      16.4612(0.025)  \\
2005hk   &      53687.1   &      16.6917(0.018)   &      16.7027(0.038)  \\
2005hk   &      53698.1   &      18.5853(0.037)   &      18.5868(0.020)  \\
2005hk   &      53699.1   &      18.9310(0.052)   &      18.7850(0.025)  \\
2005hk   &      53702.1   &      19.3411(0.064)   &      19.2887(0.041)  \\
\enddata
\tablecomments{$\Delta S$ values used in this analysis may be calculated directly from the table data by taking the difference of the CSP and SDSS magnitudes. To account for spectrum-template mismatch uncertainties, an extra uncertainty of 0.038 magnitudes should be combined in quadrature with the photometric uncertainties given here. }
\end{deluxetable}
\end{center}

 \clearpage


\begin{center}
\begin{deluxetable}{ccccc}
\tablewidth{0pt}
\tabletypesize{\tiny}
\tablecaption{Magnitude data - $g$ band  \label{table:gmagdata}}
\tablehead{
        \colhead{IAU} &
        \colhead{MJD}&
        \colhead{CSP(SDSS)} &
        \colhead{SDSS(native)} &
        \colhead{spectrum(1=yes)} 
	}
\startdata
2005hc   &      53663.3   &      17.3418(0.007)   &      17.3340(0.013)   &   0  \\
2005hc   &      53664.4   &      17.3041(0.007)   &      17.3210(0.007)   &   0  \\
2005hc   &      53665.4   &      17.2714(0.006)   &      17.2780(0.020)   &   0  \\
2005hc   &      53666.4   &      17.2490(0.008)   &      17.2920(0.014)   &   0  \\
2005hc   &      53668.3   &      17.2740(0.006)   &      17.2590(0.035)   &   0  \\
2005hc   &      53669.3   &      17.3012(0.005)   &      17.3130(0.015)   &   0  \\
2005hc   &      53669.4   &      17.3040(0.005)   &      17.3140(0.010)   &   0  \\
2005hc   &      53670.3   &      17.3283(0.006)   &      17.3110(0.014)   &   0  \\
2005hc   &      53671.4   &      17.3610(0.007)   &      17.3740(0.011)   &   0  \\
2005hc   &      53673.3   &      17.4444(0.008)   &      17.4290(0.013)   &   0  \\
2005hc   &      53674.3   &      17.4975(0.006)   &      17.5120(0.012)   &   0  \\
2005hc   &      53675.3   &      17.5558(0.008)   &      17.5610(0.011)   &   0  \\
2005hc   &      53676.4   &      17.6242(0.012)   &      17.6090(0.016)   &   0  \\
2005hc   &      53680.3   &      17.9094(0.040)   &      17.9380(0.013)   &   0  \\
2005hc   &      53681.4   &      17.9970(0.039)   &      18.0150(0.007)   &   0  \\
2005hc   &      53684.3   &      18.2722(0.013)   &      18.3430(0.026)   &   0  \\
2005hc   &      53686.3   &      18.4494(0.019)   &      18.4600(0.015)   &   0  \\
2005hc   &      53687.4   &      18.5501(0.022)   &      18.4660(0.047)   &   0  \\
2005hc   &      53693.3   &      19.0469(0.021)   &      19.0660(0.024)   &   0  \\
2005hc   &      53697.3   &      19.3663(0.014)   &      19.3460(0.019)   &   0  \\
2005hc   &      53698.3   &      19.4643(0.014)   &      19.4570(0.021)   &   0  \\
2005hc   &      53700.3   &      19.6331(0.013)   &      19.5760(0.017)   &   0  \\
2005hc   &      53704.3   &      19.8622(0.020)   &      19.7980(0.020)   &   0  \\
2005ir   &      53682.1   &      18.4005(0.016)   &      18.3917(0.016)   &   0  \\
2005ir   &      53684.2   &      18.3855(0.013)   &      18.3708(0.014)   &   0  \\
2005ir   &      53694.2   &      18.7716(0.018)   &      18.7213(0.022)   &   0  \\
2005ir   &      53695.2   &      18.9059(0.024)   &      18.7907(0.018)   &   0  \\
2005ir   &      53698.1   &      19.0582(0.014)   &      19.0233(0.014)   &   0  \\
2005ir   &      53703.2   &      19.5733(0.023)   &      19.4549(0.014)   &   0  \\
2005ku   &      53699.1   &      17.6318(0.013)   &      17.5850(0.030)   &   0  \\
2007jg   &      54376.5   &      18.0142(0.006)   &      18.0340(0.023)   &   0  \\
2007jg   &      54382.5   &      18.6406(0.011)   &      18.6820(0.018)   &   0  \\
2007jg   &      54385.5   &      18.9678(0.013)   &      18.9700(0.018)   &   0  \\
2007jg   &      54391.5   &      19.5170(0.031)   &      19.4880(0.026)   &   0  \\
2007jg   &      54394.5   &      19.7368(0.022)   &      19.7500(0.041)   &   0  \\
2007jg   &      54396.5   &      19.8277(0.047)   &      19.9100(0.035)   &   0  \\
2007jg   &      54411.3   &      20.5750(0.033)   &      20.5990(0.030)   &   0  \\
2007jg   &      54413.4   &      20.6624(0.059)   &      20.5600(0.047)   &   0  \\
2007jg   &      54415.4   &      20.6805(0.046)   &      20.5990(0.027)   &   0  \\
2007jg   &      54418.4   &      20.6964(0.059)   &      20.6800(0.039)   &   0  \\
2007jg   &      54422.4   &      20.7014(0.052)   &      20.6830(0.034)   &   0  \\
2007jh   &      54364.4   &      18.4714(0.011)   &      18.5278(0.008)   &   0  \\
2007jh   &      54380.4   &      20.0991(0.026)   &      19.9968(0.035)   &   0  \\
2007jh   &      54392.4   &      20.9677(0.065)   &      21.0601(0.047)   &   0  \\
2007mm   &      54385.2   &      20.2828(0.028)   &      20.4044(0.033)   &   0  \\
2007mm   &      54392.2   &      19.6799(0.015)   &      19.7425(0.030)   &   0  \\
2007mm   &      54394.2   &      19.6984(0.019)   &      19.7518(0.029)   &   0  \\
2007mm   &      54395.1   &      19.7796(0.024)   &      19.8230(0.037)   &   0  \\
2007mm   &      54409.2   &      21.5620(0.045)   &      21.6200(0.049)   &   0  \\
2005hc   &      53665.0   &      17.2857(0.007)   &      17.2912(0.015)   &   1  \\
2005hc   &      53667.0   &      17.2582(0.009)   &      17.2821(0.025)   &   1  \\
2007jg   &      54384.0   &      18.8091(0.009)   &      18.8353(0.016)   &   1  \\
2007jg   &      54389.0   &      19.2766(0.033)   &      19.2733(0.024)   &   1  \\
2007jg   &      54393.0   &      19.6526(0.021)   &      19.6252(0.032)   &   1  \\
2007jh   &      54380.4   &      20.0876(0.027)   &      19.9969(0.034)   &   1  \\
2005gj   &      53699.0   &      17.8867(0.010)   &      17.9020(0.011)   &   1  \\
2005hk   &      53677.2   &      16.3364(0.006)   &      16.3616(0.005)   &   1  \\
2005hk   &      53678.2   &      16.1808(0.007)   &      16.1949(0.008)   &   1  \\
2005hk   &      53680.3   &      15.9741(0.011)   &      15.9594(0.006)   &   1  \\
2005hk   &      53684.0   &      15.7797(0.006)   &      15.7907(0.012)   &   1  \\
2005hk   &      53697.2   &      16.7410(0.008)   &      16.7724(0.016)   &   1  \\
2005hk   &      53699.1   &      16.9971(0.006)   &      16.9940(0.015)   &   1  \\
\enddata
\tablecomments{$\Delta S$ values used in this analysis may be calculated by taking the difference of the CSP and SDSS magnitudes, and combining an extra 0.013 magnitudes in quadrature with the given uncertainties, to account for template-spectrum mismatch uncertainties.}
\end{deluxetable}
\end{center}

 \clearpage


\begin{center}
\begin{deluxetable}{ccccc}
\tablewidth{0pt}
\tabletypesize{\tiny}
\tablecaption{Magnitude data - $r$ band\label{table:rmagdata}}
\tablehead{
        \colhead{IAU} &
        \colhead{MJD}&
        \colhead{CSP(SDSS)} &
        \colhead{SDSS(native)} &
        \colhead{spectrum(1=yes)}
        }
\startdata
2005hc   &      53663.3   &      17.4912(0.010)   &      17.4990(0.010)   &   0  \\
2005hc   &      53664.4   &      17.4371(0.009)   &      17.4490(0.013)   &   0  \\
2005hc   &      53665.4   &      17.3927(0.007)   &      17.4760(0.022)   &   0  \\
2005hc   &      53666.4   &      17.3672(0.008)   &      17.3530(0.013)   &   0  \\
2005hc   &      53668.3   &      17.3506(0.007)   &      17.3530(0.023)   &   0  \\
2005hc   &      53669.3   &      17.3483(0.006)   &      17.3970(0.012)   &   0  \\
2005hc   &      53669.4   &      17.3480(0.006)   &      17.3520(0.016)   &   0  \\
2005hc   &      53673.3   &      17.4282(0.009)   &      17.4120(0.007)   &   0  \\
2005hc   &      53674.3   &      17.4665(0.007)   &      17.5010(0.013)   &   0  \\
2005hc   &      53675.3   &      17.5063(0.009)   &      17.5080(0.025)   &   0  \\
2005hc   &      53676.4   &      17.5545(0.013)   &      17.6140(0.008)   &   0  \\
2005hc   &      53680.3   &      17.7597(0.025)   &      17.7840(0.007)   &   0  \\
2005hc   &      53681.4   &      17.8203(0.024)   &      17.8470(0.006)   &   0  \\
2005hc   &      53684.3   &      17.9885(0.011)   &      18.0000(0.018)   &   0  \\
2005hc   &      53686.3   &      18.0456(0.014)   &      18.0570(0.012)   &   0  \\
2005hc   &      53687.4   &      18.0594(0.016)   &      18.0010(0.039)   &   0  \\
2005hc   &      53693.3   &      18.1736(0.013)   &      18.2090(0.028)   &   0  \\
2005hc   &      53697.3   &      18.3245(0.009)   &      18.3420(0.019)   &   0  \\
2005hc   &      53698.3   &      18.3672(0.010)   &      18.3740(0.029)   &   0  \\
2005hc   &      53700.3   &      18.4720(0.010)   &      18.4980(0.012)   &   0  \\
2005hc   &      53704.3   &      18.7164(0.014)   &      18.7360(0.019)   &   0  \\
2005ir   &      53682.1   &      18.5120(0.022)   &      18.4856(0.013)   &   0  \\
2005ir   &      53684.2   &      18.4458(0.015)   &      18.4201(0.013)   &   0  \\
2005ir   &      53694.2   &      18.5608(0.015)   &      18.5783(0.017)   &   0  \\
2005ir   &      53695.2   &      18.6981(0.021)   &      18.6293(0.015)   &   0  \\
2005ir   &      53698.1   &      18.7881(0.016)   &      18.7951(0.013)   &   0  \\
2005ir   &      53703.2   &      19.0699(0.020)   &      19.0249(0.017)   &   0  \\
2005ku   &      53699.1   &      17.5999(0.016)   &      17.5379(0.009)   &   0  \\
2007jg   &      54376.5   &      17.8088(0.009)   &      17.8150(0.009)   &   0  \\
2007jg   &      54382.5   &      18.1906(0.015)   &      18.1710(0.017)   &   0  \\
2007jg   &      54385.5   &      18.2596(0.017)   &      18.2350(0.020)   &   0  \\
2007jg   &      54391.5   &      18.3493(0.022)   &      18.3900(0.021)   &   0  \\
2007jg   &      54394.5   &      18.5428(0.018)   &      18.5380(0.022)   &   0  \\
2007jg   &      54396.5   &      18.6699(0.018)   &      18.7020(0.021)   &   0  \\
2007jg   &      54402.5   &      19.0779(0.024)   &      19.1280(0.029)   &   0  \\
2007jg   &      54411.3   &      19.5144(0.035)   &      19.4590(0.031)   &   0  \\
2007jg   &      54415.4   &      19.6723(0.042)   &      19.4860(0.017)   &   0  \\
2007jg   &      54417.4   &      19.7573(0.039)   &      19.6040(0.044)   &   0  \\
2007jg   &      54418.4   &      19.8032(0.037)   &      19.6570(0.027)   &   0  \\
2007jg   &      54422.4   &      19.8705(0.040)   &      19.7610(0.030)   &   0  \\
2007jg   &      54424.3   &      19.8735(0.054)   &      19.8340(0.028)   &   0  \\
2007jh   &      54364.4   &      18.3229(0.016)   &      18.2965(0.008)   &   0  \\
2007jh   &      54380.4   &      18.8176(0.035)   &      18.8008(0.024)   &   0  \\
2007jh   &      54392.4   &      19.6852(0.042)   &      19.7335(0.031)   &   0  \\
2007jh   &      54394.3   &      19.7882(0.046)   &      19.7882(0.039)   &   0  \\
2007jh   &      54395.4   &      19.9599(0.046)   &      19.8144(0.040)   &   0  \\
2007mm   &      54385.2   &      20.1710(0.039)   &      20.2331(0.026)   &   0  \\
2007mm   &      54392.2   &      19.3080(0.016)   &      19.3185(0.045)   &   0  \\
2007mm   &      54394.2   &      19.2470(0.019)   &      19.3214(0.030)   &   0  \\
2007mm   &      54395.1   &      19.1608(0.020)   &      19.3478(0.039)   &   0  \\
2007mm   &      54400.2   &      19.4571(0.054)   &      19.5616(0.049)   &   0  \\
2007mm   &      54403.2   &      19.7147(0.022)   &      19.7641(0.047)   &   0  \\
2007mm   &      54409.2   &      20.2400(0.023)   &      20.2895(0.023)   &   0  \\
2007mm   &      54418.2   &      20.9156(0.056)   &      20.9182(0.047)   &   0  \\
2005hc   &      53701.0   &      18.5340(0.010)   &      18.5478(0.017)   &   1  \\
2007jg   &      54389.0   &      18.2992(0.028)   &      18.2892(0.021)   &   1  \\
2007jh   &      54380.4   &      18.8293(0.031)   &      18.8028(0.023)   &   1  \\
2005gj   &      53699.0   &      17.2536(0.009)   &      17.2639(0.006)   &   1  \\
2005hk   &      53678.2   &      16.2680(0.007)   &      16.2907(0.008)   &   1  \\
2005hk   &      53680.3   &      16.0621(0.012)   &      16.0522(0.005)   &   1  \\
2005hk   &      53684.0   &      15.8228(0.008)   &      15.8036(0.011)   &   1  \\
\enddata
\tablecomments{$\Delta S$ values used in this analysis may be calculated by taking the difference of the CSP and SDSS magnitudes, and combining an extra 0.009 magnitudes in quadrature with the given uncertainties, to account for template-spectrum mismatch uncertainties.}
\end{deluxetable}
\end{center}

 \clearpage
 

\begin{center}
\begin{deluxetable}{ccccc}
\tablewidth{0pt}
\tabletypesize{\tiny}
\tablecaption{Magnitude data - $i$ band\label{table:imagdata}}
\tablehead{
        \colhead{IAU} &
        \colhead{MJD}&
        \colhead{CSP(SDSS)} &
        \colhead{SDSS(native)} &
        \colhead{spectrum(1=yes)}
        }
\startdata
2005hc   &      53664.4   &      17.9096(0.011)   &      17.8780(0.019)   &   0  \\
2005hc   &      53665.4   &      17.9195(0.009)   &      17.9680(0.015)   &   0  \\
2005hc   &      53666.4   &      17.9015(0.012)   &      17.9340(0.011)   &   0  \\
2005hc   &      53668.3   &      17.9325(0.010)   &      17.9770(0.020)   &   0  \\
2005hc   &      53669.3   &      17.9552(0.008)   &      17.9910(0.015)   &   0  \\
2005hc   &      53669.4   &      17.9568(0.008)   &      17.9530(0.018)   &   0  \\
2005hc   &      53670.3   &      17.9669(0.011)   &      18.0080(0.024)   &   0  \\
2005hc   &      53671.4   &      17.9781(0.013)   &      18.0410(0.021)   &   0  \\
2005hc   &      53673.3   &      18.0288(0.014)   &      18.0530(0.010)   &   0  \\
2005hc   &      53674.3   &      18.0753(0.012)   &      18.0990(0.012)   &   0  \\
2005hc   &      53675.3   &      18.1325(0.014)   &      18.1450(0.014)   &   0  \\
2005hc   &      53676.4   &      18.2021(0.026)   &      18.2490(0.011)   &   0  \\
2005hc   &      53684.3   &      18.6690(0.016)   &      18.7390(0.028)   &   0  \\
2005hc   &      53686.3   &      18.6669(0.022)   &      18.6570(0.027)   &   0  \\
2005hc   &      53693.3   &      18.5743(0.016)   &      18.5600(0.020)   &   0  \\
2005hc   &      53697.3   &      18.5252(0.013)   &      18.6170(0.021)   &   0  \\
2005hc   &      53698.3   &      18.5173(0.016)   &      18.5080(0.038)   &   0  \\
2005hc   &      53700.3   &      18.5578(0.016)   &      18.6230(0.017)   &   0  \\
2005hc   &      53704.3   &      18.7922(0.019)   &      18.7630(0.025)   &   0  \\
2005ir   &      53682.1   &      18.8816(0.049)   &      18.8316(0.024)   &   0  \\
2005ir   &      53684.2   &      18.7843(0.023)   &      18.8533(0.024)   &   0  \\
2005ir   &      53694.2   &      19.1188(0.030)   &      19.0689(0.019)   &   0  \\
2005ir   &      53695.2   &      19.1819(0.025)   &      19.1578(0.018)   &   0  \\
2005ir   &      53698.1   &      19.4412(0.038)   &      19.4467(0.026)   &   0  \\
2005ku   &      53699.1   &      18.0034(0.022)   &      17.9897(0.024)   &   0  \\
2007jg   &      54363.3   &      17.9352(0.017)   &      17.9743(0.020)   &   0  \\
2007jg   &      54364.4   &      17.9596(0.021)   &      17.9978(0.020)   &   0  \\
2007jg   &      54376.3   &      18.5998(0.026)   &      18.5276(0.012)   &   0  \\
2007jg   &      54378.4   &      18.7409(0.033)   &      18.6534(0.012)   &   0  \\
2007jg   &      54383.3   &      18.8369(0.037)   &      18.7547(0.012)   &   0  \\
2007jg   &      54385.4   &      18.8559(0.040)   &      18.6956(0.012)   &   0  \\
2007jg   &      54392.3   &      18.6087(0.028)   &      18.5657(0.033)   &   0  \\
2007jg   &      54397.2   &      18.7282(0.026)   &      18.7050(0.032)   &   0  \\
2007jg   &      54403.3   &      19.2684(0.039)   &      19.1936(0.032)   &   0  \\
2007jg   &      54411.3   &      19.7462(0.054)   &      19.6081(0.025)   &   0  \\
2007jh   &      54364.4   &      18.5514(0.034)   &      18.4944(0.013)   &   0  \\
2007jh   &      54380.4   &      18.8483(0.023)   &      18.7984(0.020)   &   0  \\
2007jh   &      54392.4   &      19.6366(0.059)   &      19.5835(0.031)   &   0  \\
2007jh   &      54394.3   &      19.7361(0.046)   &      19.6203(0.045)   &   0  \\
2007jh   &      54395.4   &      19.7462(0.062)   &      19.6713(0.044)   &   0  \\
2007mm   &      54385.2   &      20.1100(0.057)   &      20.2173(0.030)   &   0  \\
2007mm   &      54392.2   &      19.4221(0.018)   &      19.5101(0.038)   &   0  \\
2007mm   &      54394.2   &      19.3601(0.021)   &      19.4240(0.036)   &   0  \\
2007mm   &      54395.1   &      19.3341(0.025)   &      19.4055(0.044)   &   0  \\
2007mm   &      54400.2   &      19.4877(0.041)   &      19.5024(0.036)   &   0  \\
2007mm   &      54403.2   &      19.6725(0.024)   &      19.6163(0.032)   &   0  \\
2007mm   &      54409.2   &      19.9729(0.027)   &      19.9563(0.024)   &   0  \\
2005hc   &      53701.0   &      18.5667(0.014)   &      18.6656(0.023)   &   1  \\
2007jg   &      54389.0   &      18.5739(0.053)   &      18.6181(0.022)   &   1  \\
2005hj   &      53675.0   &      18.3083(0.013)   &      18.3105(0.009)   &   1  \\
2005hj   &      53677.0   &      18.3595(0.026)   &      18.3486(0.012)   &   1  \\
2005hj   &      53678.0   &      18.3716(0.032)   &      18.3640(0.019)   &   1  \\
2005hj   &      53680.0   &      18.4047(0.042)   &      18.3763(0.019)   &   1  \\
2005hj   &      53685.0   &      18.5242(0.044)   &      18.6328(0.015)   &   1  \\
2005hj   &      53700.0   &      18.9963(0.021)   &      18.9838(0.022)   &   1  \\
2005hk   &      53678.2   &      16.4768(0.008)   &      16.5281(0.013)   &   1  \\
2005hk   &      53680.3   &      16.2838(0.012)   &      16.2988(0.009)   &   1  \\
2005hk   &      53684.0   &      16.0581(0.009)   &      16.0595(0.011)   &   1  \\
2005hk   &      53699.1   &      15.8532(0.006)   &      15.8586(0.010)   &   1  \\
\enddata
\tablecomments{$\Delta S$ values used in this analysis may be calculated by taking the difference of the CSP and SDSS magnitudes, and combining an extra 0.014 magnitudes in quadrature with the given uncertainties, to account for template-spectrum mismatch uncertainties.}
\tablecomments{See comments forTable~\ref{table:gmagdata}.}
\end{deluxetable}
\end{center}

 \clearpage

\clearpage
\begin{figure}
\figurenum{1}
\epsscale{1.0}
\plottwo{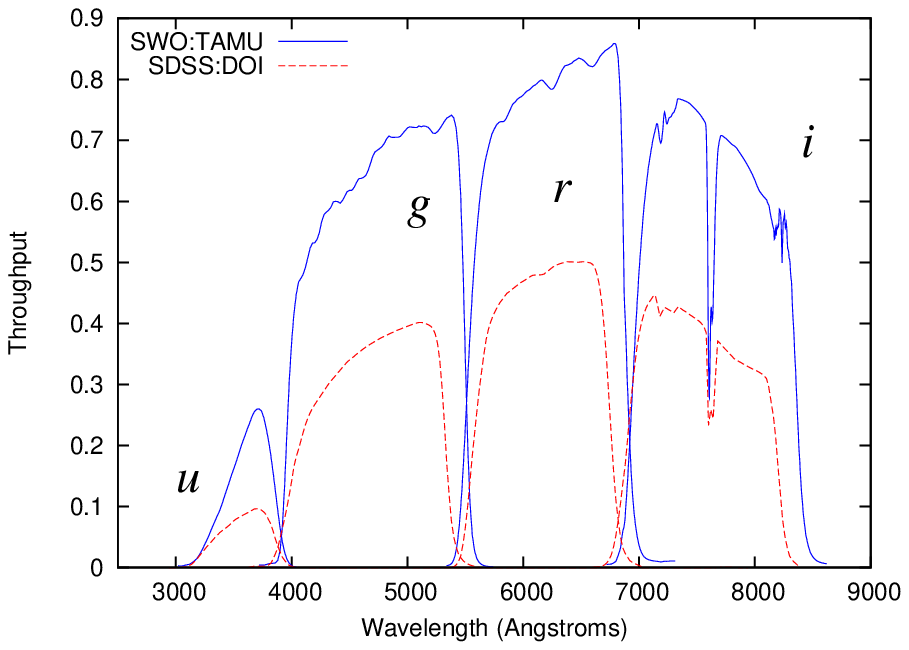}{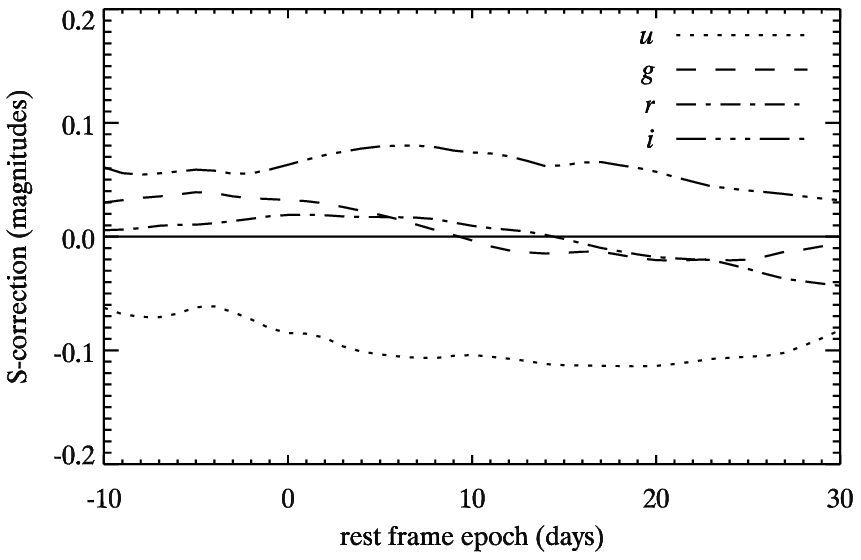} 
\caption[scorr_fig.eps]{ CSP and SDSS throughputs are plotted in the left panel. The right panel shows $S$-corrections for a mean SN~Ia observed at a redshift of 0.04 (the average redshift of the SNe in our sample) as a function of time. A solid line has been drawn at $S$-correction equals zero to guide the eye. Descriptions of the filter response functions used can be found in \citet{Doi:2010} and \citet{Max:2011}. Hsiao SN~Ia templates have been used as a proxy for the mean SN~Ia spectral energy distributions. }
\label{fig:scorr_mags}
\end{figure}

\clearpage
\begin{figure}
\figurenum{2}
\epsscale{1.0}
\includegraphics[angle=+90, scale=0.75]{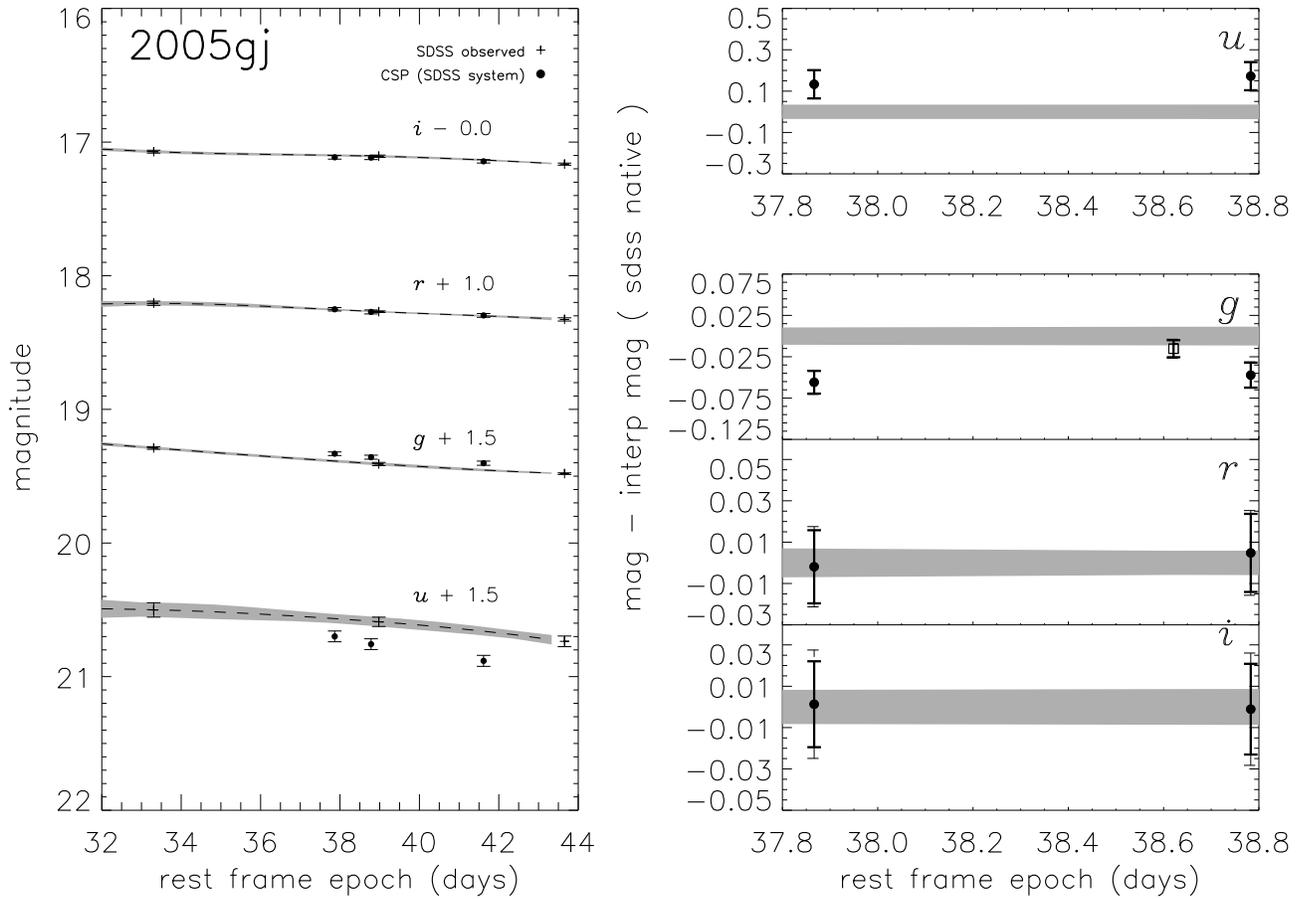}
\caption[fig2.eps]{SN4524 / {\SNgj} : SDSS native photometry, $S$-corrected CSP  photometry and spline fits to the SDSS photometry are shown in the left panel. The right panels show $\Delta m$, defined as $S$-corrected CSP magnitude minus interpolated SDSS magnitude plotted as a function of rest frame epoch. The shaded gray bars show the uncertainty in the interpolated SDSS magnitude. The $u$-band data are included when available. \label{fig:SN4524}}
\end{figure}

\clearpage
\begin{figure}
\figurenum{3}
\epsscale{1.0}
\includegraphics[angle=+90, scale=0.75]{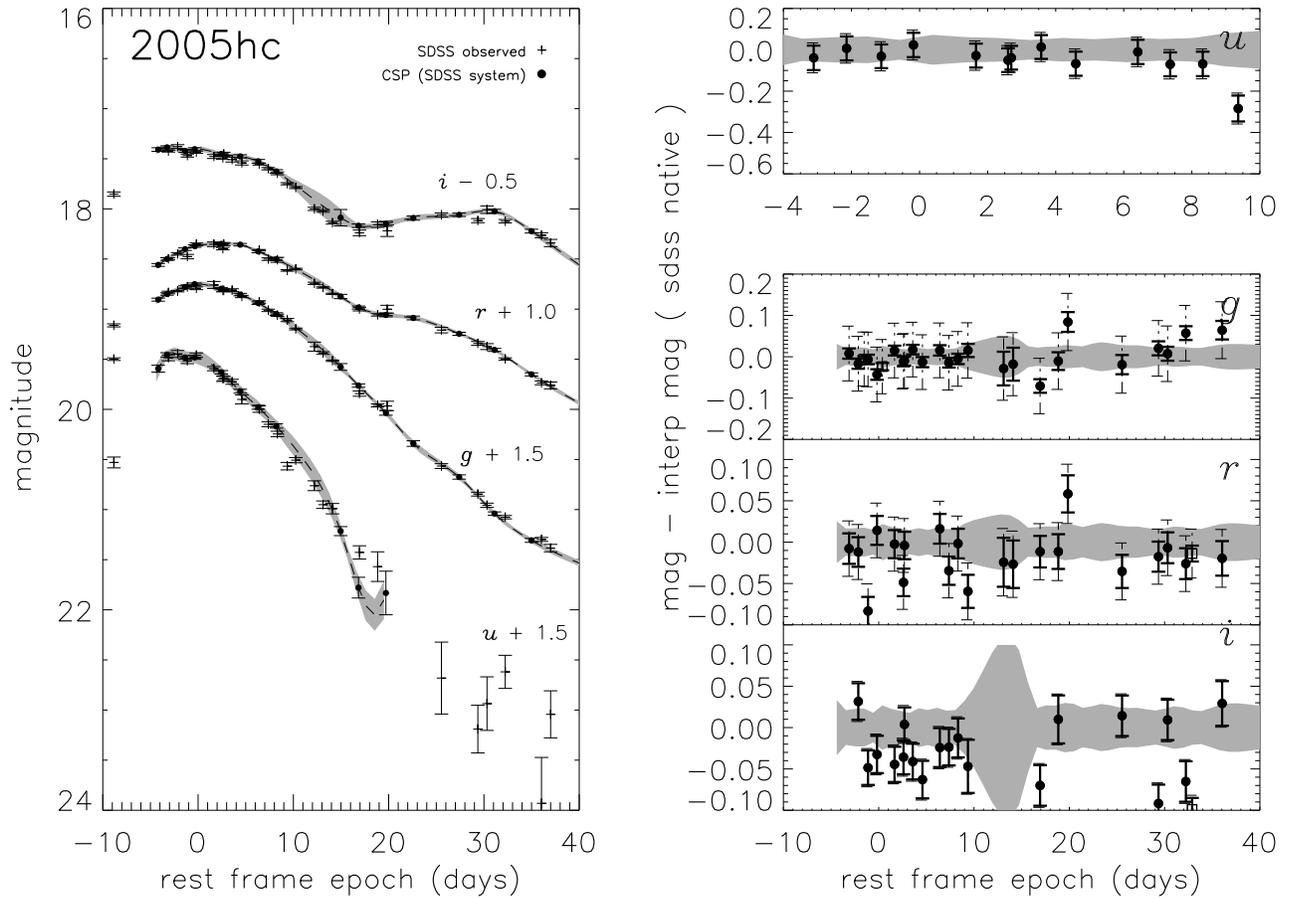}
\caption[fig3.eps]{SN5944 / {\SNhc} : SDSS native photometry, $S$-corrected CSP  photometry and spline fits to the CSP photometry are shown in the left panel. The right panels show $\Delta m$, defined as interpolated $S$-corrected CSP magnitude $-$ SDSS magnitude plotted as a function of rest frame epoch. The shaded gray bars show the uncertainty in the SDSS magnitudes.  \label{fig:SN5944}}
\end{figure}

\clearpage
\begin{figure}
\figurenum{4}
\epsscale{1.0}
\includegraphics[angle=+90, scale=0.75]{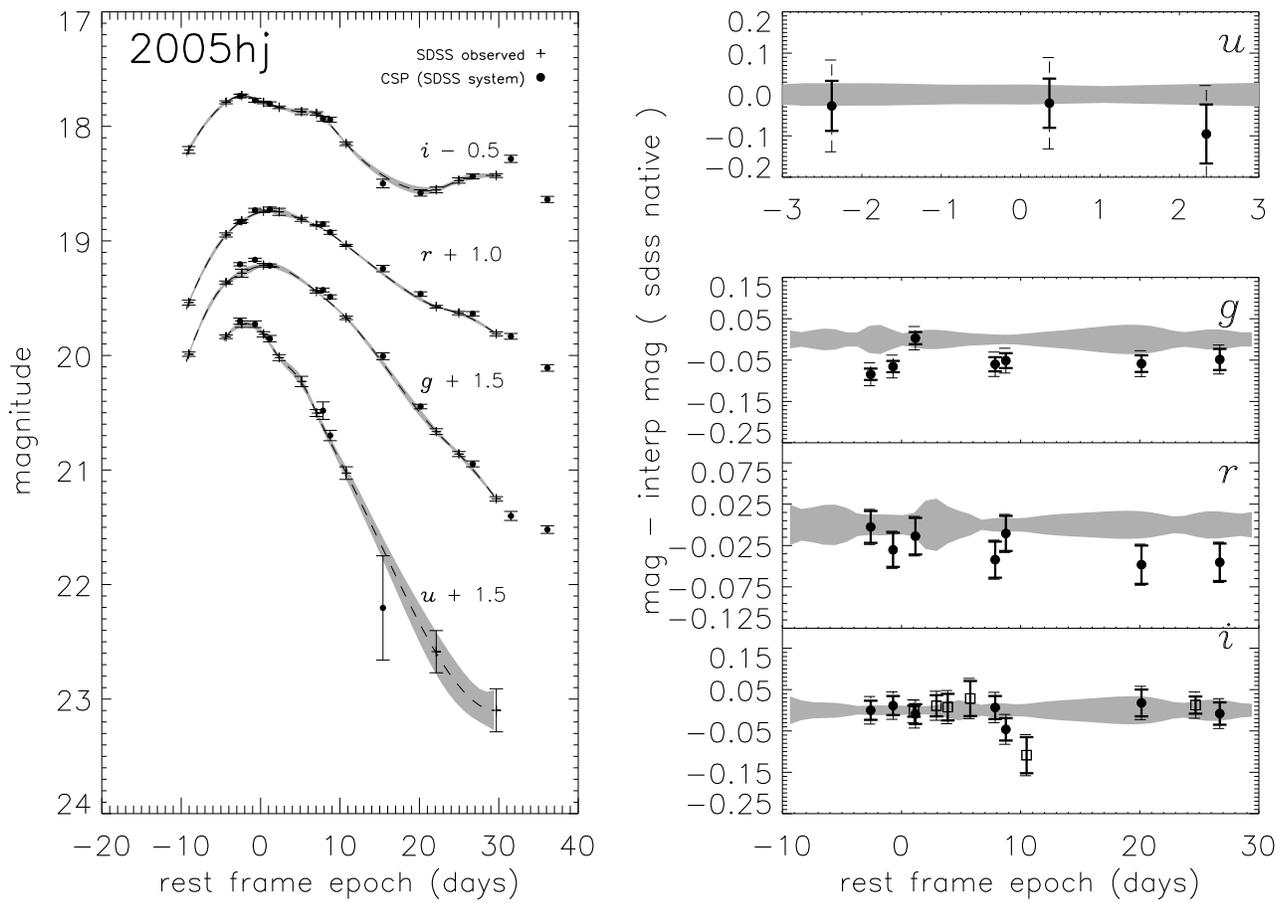}
\caption[fig4.eps]{SN6558 / {\SNhj} : quantities plotted are as described in Figure~\ref{fig:SN4524}.  \label{fig:SN6558}}
\end{figure}

\clearpage
\begin{figure}
\figurenum{5}
\epsscale{1.0}
\includegraphics[angle=+90, scale=0.75]{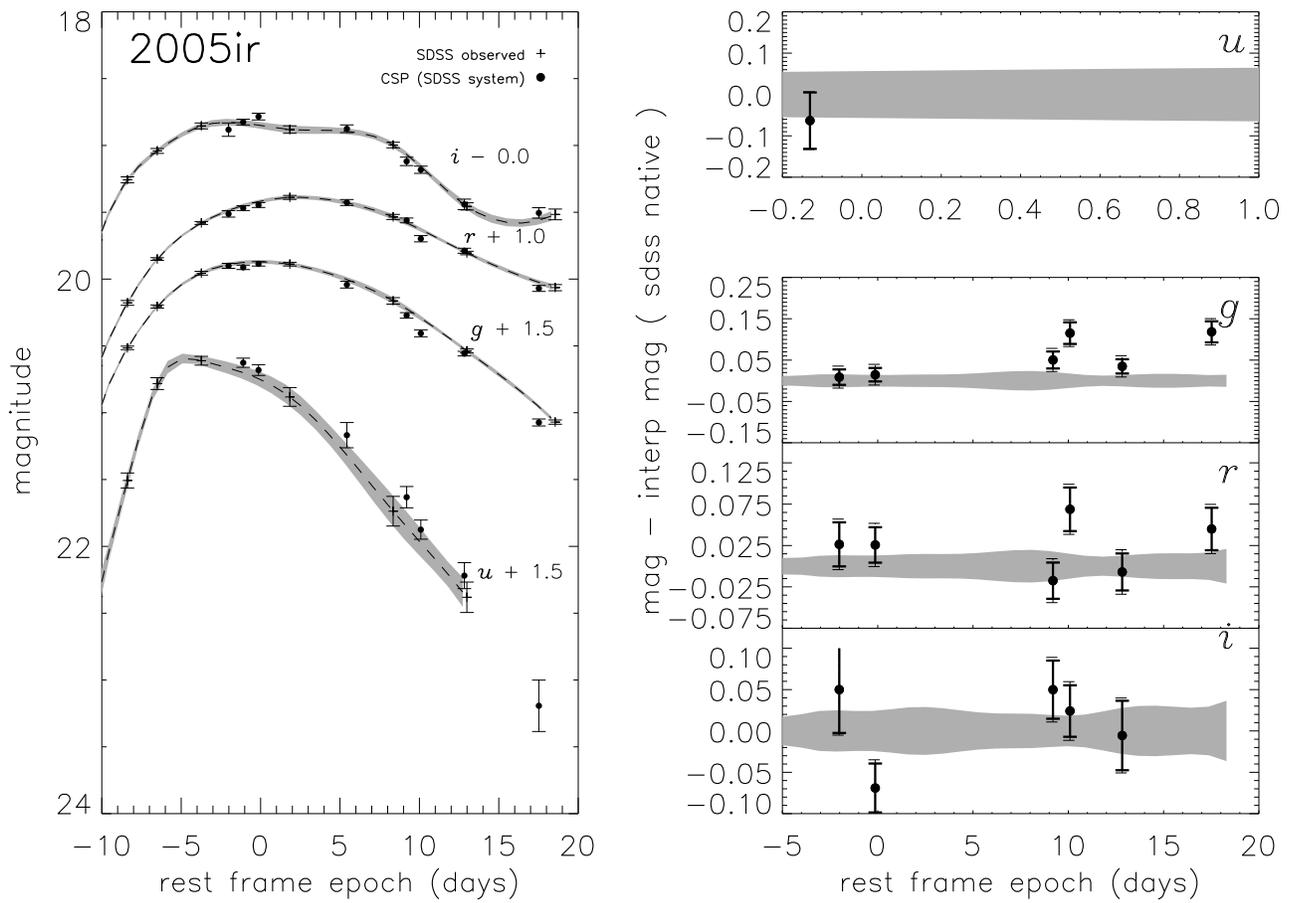}
\caption[fig5.eps]{SN7876 / {\SNir} : quantities plotted are as described in Figure~\ref{fig:SN4524}.  \label{fig:SN7876}}
\end{figure}

\clearpage
\begin{figure}
\figurenum{6}
\epsscale{1.0}
\includegraphics[angle=+90, scale=0.75]{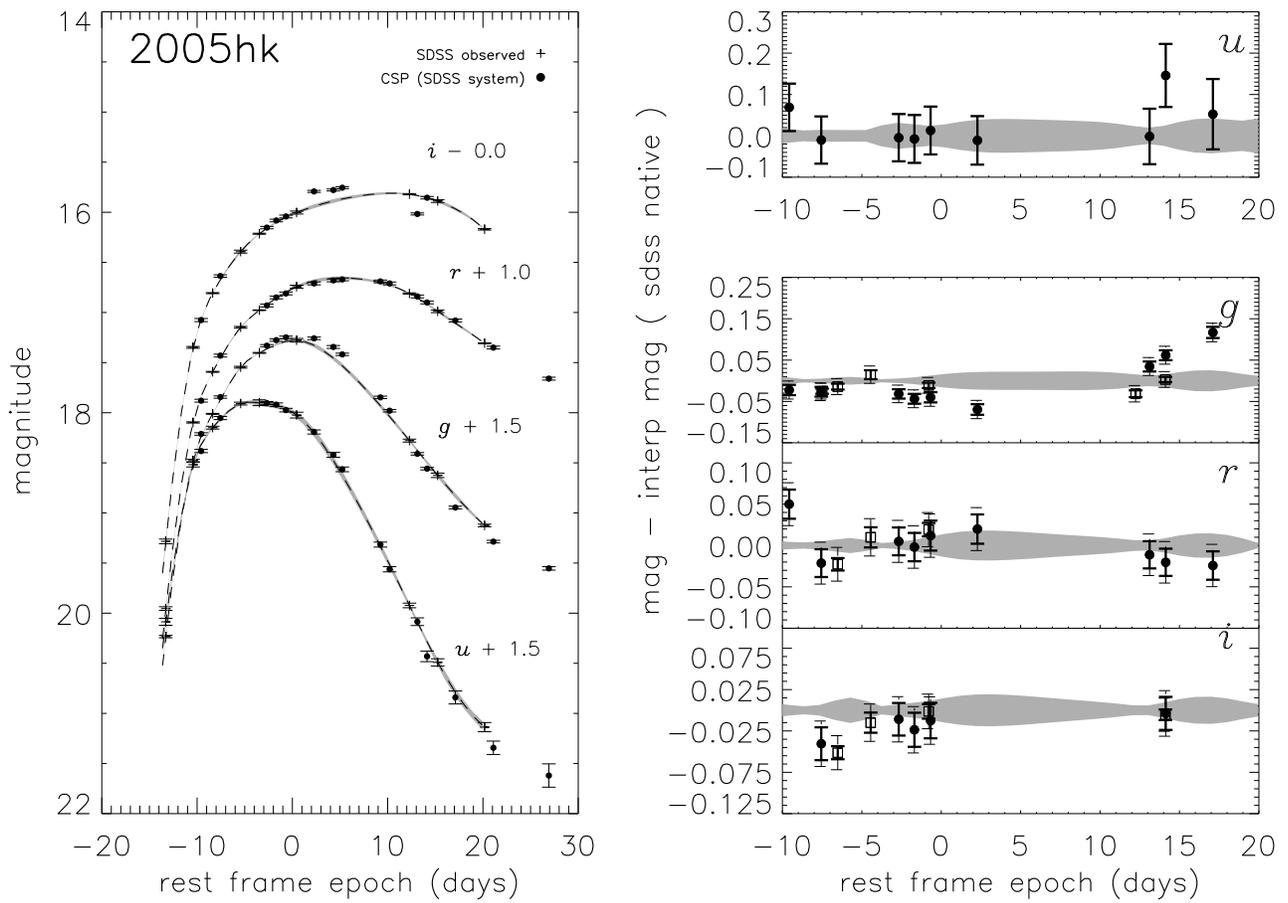}
\caption[fig6.eps]{SN8151 / {\SNhk} : quantities plotted are as described in Figure~\ref{fig:SN4524}. \label{fig:SN8151}}
\end{figure}

\clearpage
\begin{figure}
\figurenum{7}
\epsscale{1.0}
\includegraphics[angle=+90, scale=0.75]{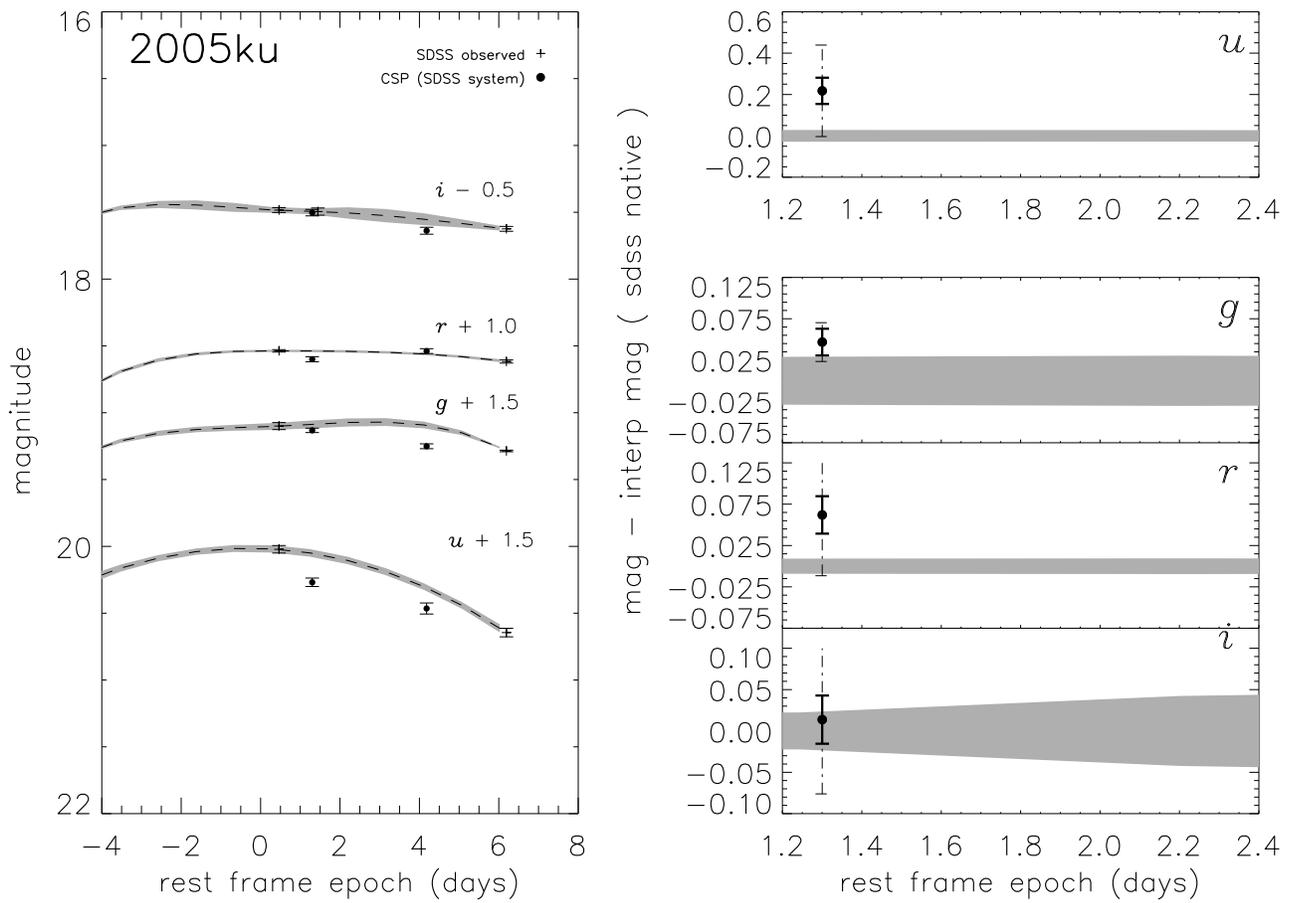}
\caption[fig7.eps]{SN10805 / {\SNku} : quantities plotted are as described in Figure~\ref{fig:SN4524}. }
\label{fig:SN10805}
\end{figure}

\clearpage
\begin{figure}
\figurenum{8}
\epsscale{1.0}
\includegraphics[angle=+90, scale=0.75]{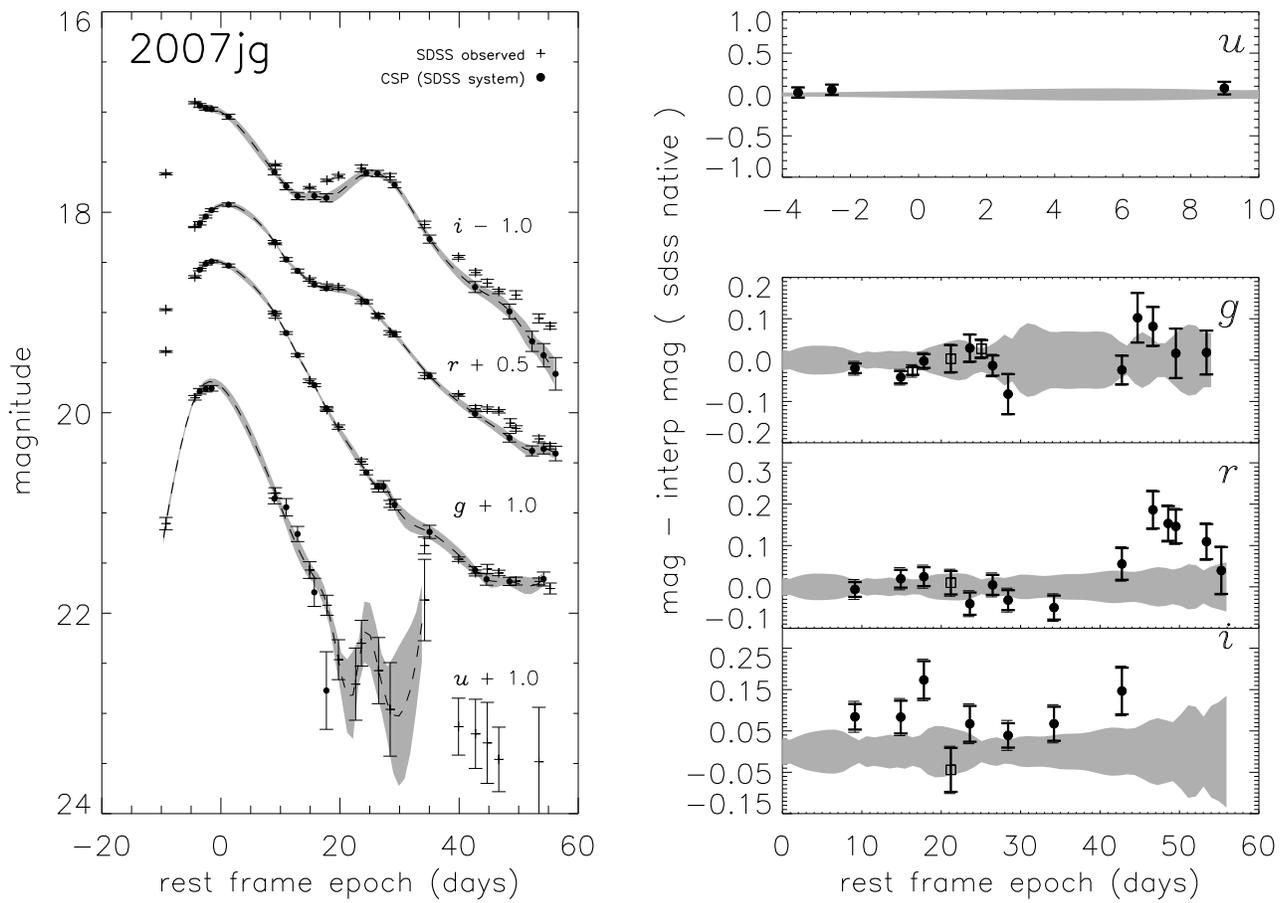}
\caption[fig8.eps]{SN17784 / {\SNjg} : quantities plotted are as described in Figure~\ref{fig:SN5944}. }	
\label{fig:SN17784}
\end{figure}

\clearpage
\begin{figure}
\figurenum{9}
\epsscale{1.0}
\includegraphics[angle=+90, scale=0.75]{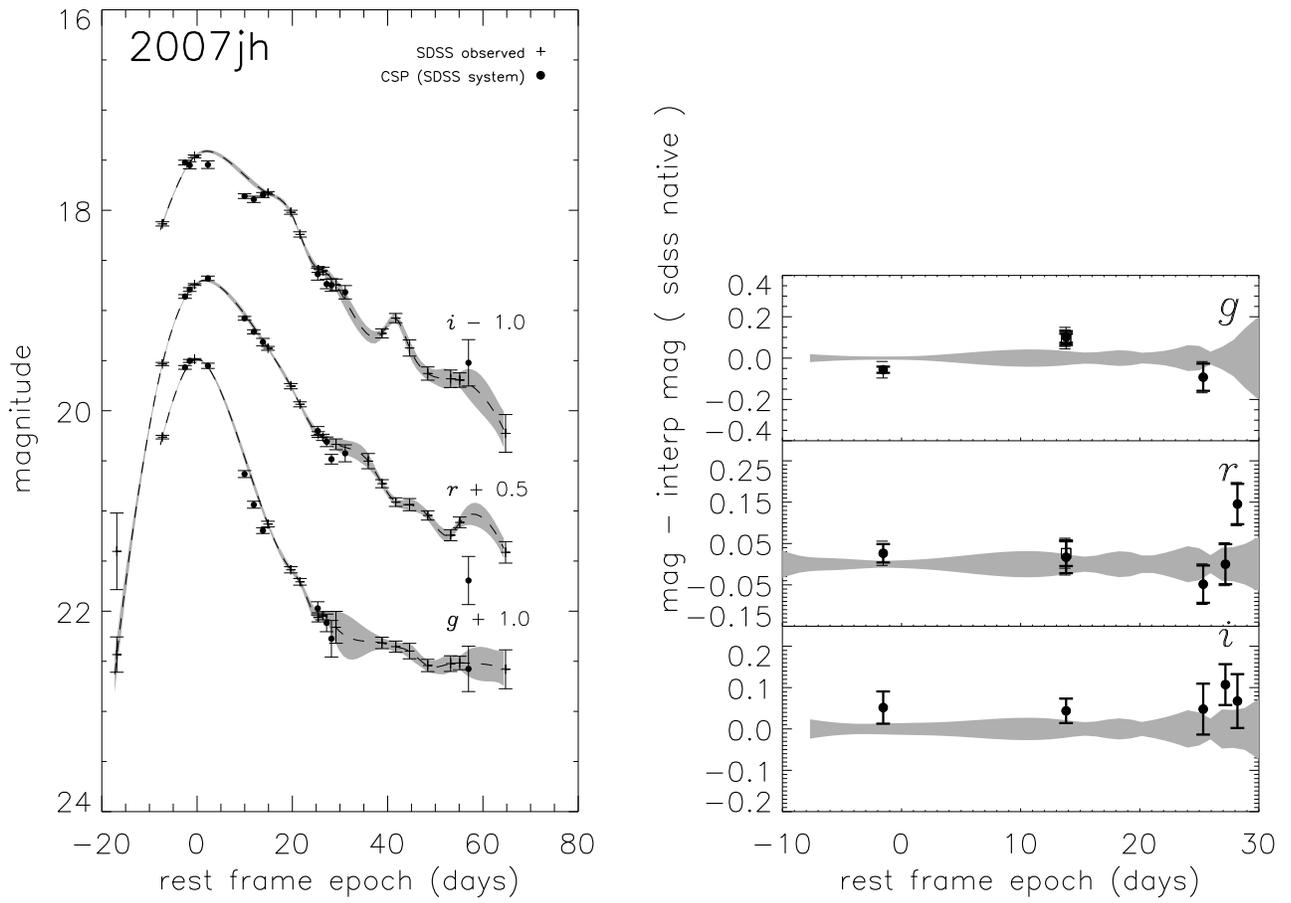}
\caption[fig9.eps]{SN17886 / {\SNjh} : quantities plotted are as described in  Figure~\ref{fig:SN4524}. }
\label{fig:SN17786}
\end{figure}

\clearpage
\begin{figure}
\figurenum{10}
\epsscale{1.0}
\includegraphics[angle=+90, scale=0.75]{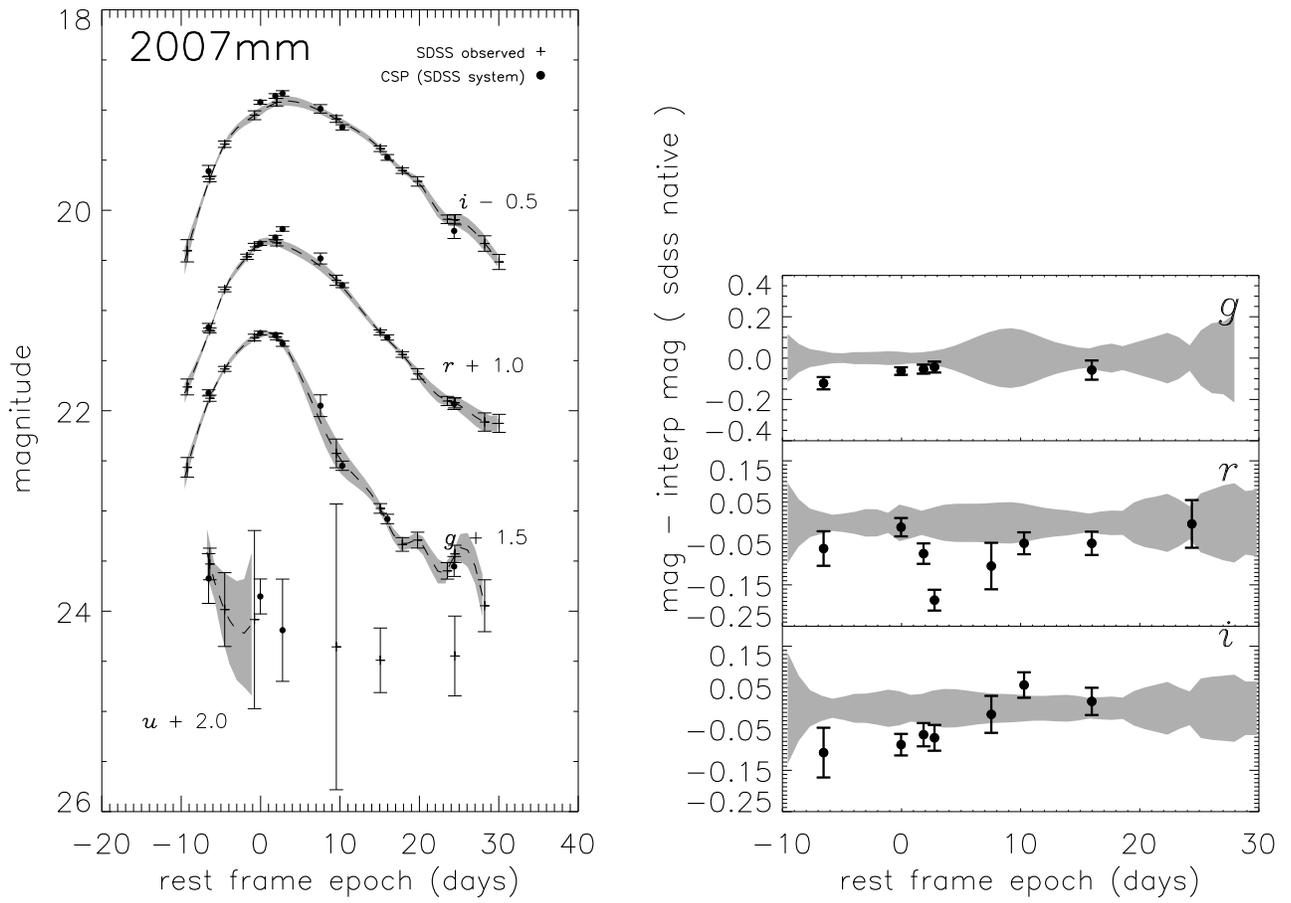}
\caption[fig10.eps]{SN18890 / {\SNmm} : quantities plotted are as described in  Figure~\ref{fig:SN4524}. }
\label{fig:SN18890}
\end{figure}

\clearpage
\begin{figure}
\figurenum{11}
\epsscale{1.0}
\includegraphics[angle=+90,scale=0.85]{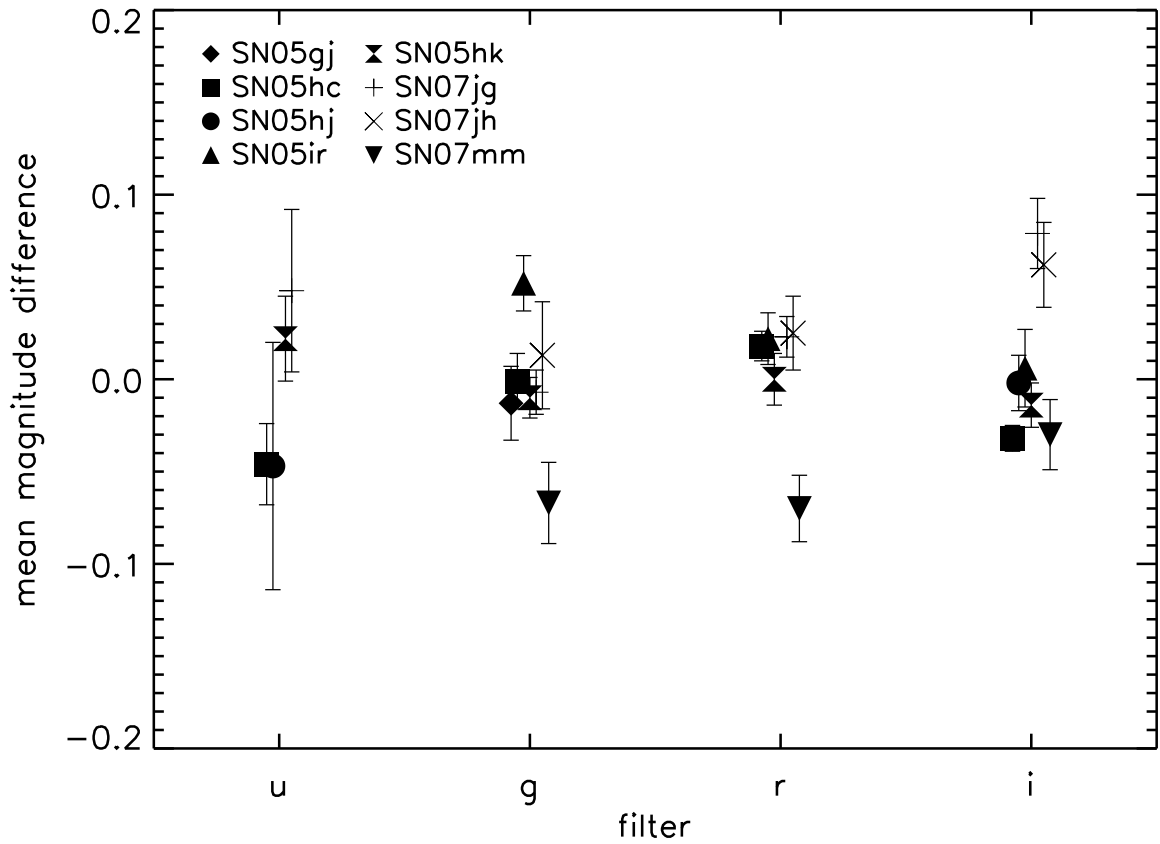}
\caption[fig11.eps]{For each supernova with at least three points in a given filter, magnitude residuals have been combined into a single mean residual and plotted as a function of filter. }
\label{fig:SNscatt}
\end{figure}
	
\clearpage
\begin{figure}
\figurenum{12}
\epsscale{1.0}
\includegraphics[scale=0.85]{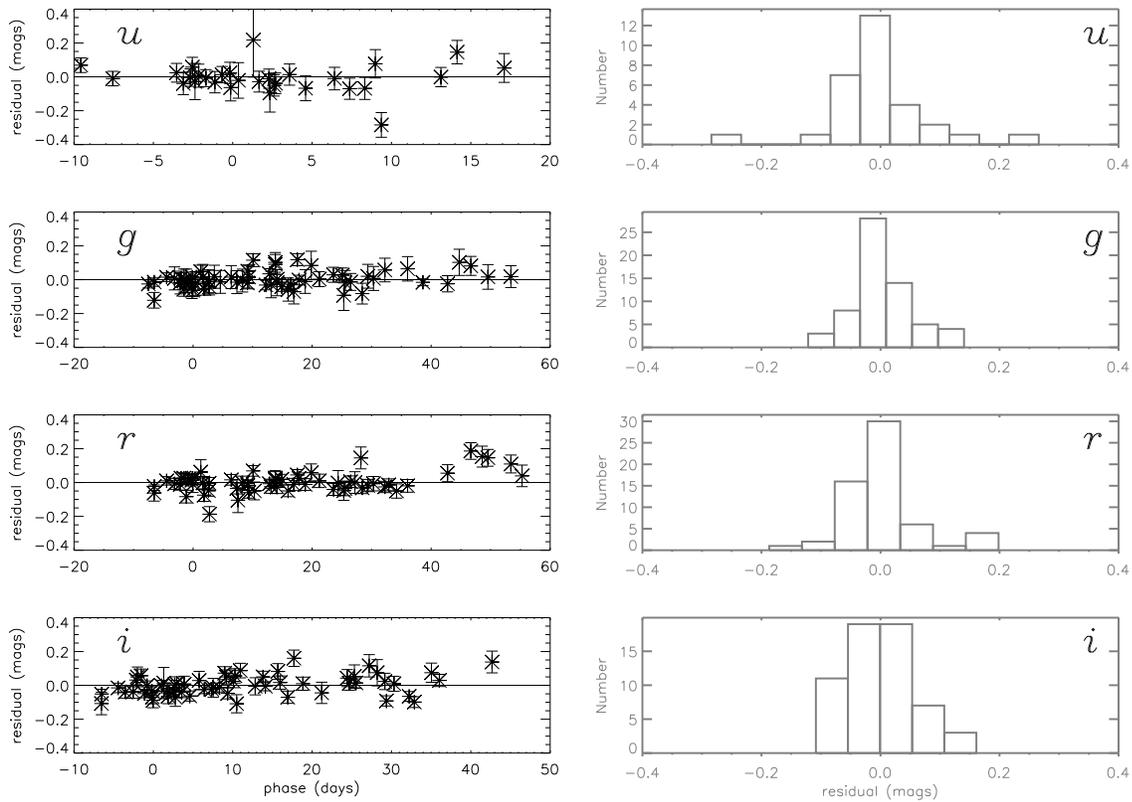}
\caption[fig12.eps]{Magnitude residuals between CSP and SDSS-II data are plotted as a function of phase, and binned into histograms. A slight trend in $i$-band residual as a function of phase is observed. Similary, $i$-band residuals do not appear to be gaussianly distributed. All other bands show minimal residual variation with phase, and reasonably gaussian distributions centered on a 0.0 magnitude difference.  }
\label{fig:phaseplot}
\end{figure}

\end{document}